\documentclass[11pt]{article}
\usepackage{latexsym}
\usepackage{amsmath}
\usepackage{amsfonts}
\usepackage{dsfont}
\usepackage{verbatim}
\usepackage{hyperref}
\usepackage{subcaption}
\usepackage{amsmath,amssymb}
\usepackage{graphicx}
\usepackage{color}
\usepackage{enumerate}
\usepackage{booktabs}
\usepackage{multirow}
\usepackage{url}
\usepackage[shortlabels]{enumitem}
\usepackage{notoccite}
\usepackage[ruled,vlined]{algorithm2e}
\usepackage{xcolor,colortbl}
\usepackage{placeins}
\usepackage{authblk}
\usepackage{array}

\title{ Inferring Cosmological Parameters with\\ Evidential Physics-Informed Neural Networks}
\author{Hai Siong Tan}

\affil{Gryphon Center for A.I. and Theoretical Sciences, Singapore}

\def\be{\begin{equation}}
\def\ee{\end{equation}}
\def\bea{\begin{eqnarray}}
\def\eea{\end{eqnarray}}

\makeatletter
\renewcommand{\@algocf@capt@plain}{above}
\makeatother

\begin{document}

\maketitle

\begin{center}
\textbf{Abstract}
\end{center}
We examine the use of a novel variant of Physics-Informed Neural Networks to predict cosmological parameters from recent supernovae and baryon acoustic oscillations (BAO) datasets. Our machine learning framework generates uncertainty estimates for target variables and the inferred unknown parameters of the underlying PDE descriptions. Built upon a hybrid of the principles of Evidential Deep Learning, Physics-Informed Neural Networks, Bayesian Neural Networks and Gaussian Processes, our model enables learning of the posterior distribution of the unknown PDE parameters through standard gradient-descent based training. We apply our model to an up-to-date BAO dataset (Bousis et al. 2024) calibrated with the CMB-inferred sound horizon, and the Pantheon+ Sne Ia distances (Scolnic et al. 2018), examining the relative effectiveness and mutual consistency among the standard $\Lambda$CDM, $w$CDM and $\Lambda_s$CDM models. Unlike previous results arising from the standard approach of minimizing an appropriate $\chi^2$ function, the posterior distributions for parameters in various models trained purely on Pantheon$+$ data were found to be largely contained within the $2\sigma$ contours of their counterparts trained on BAO data. Their posterior medians for $h_0$ were within about $2\sigma$ of one another, indicating that our machine-learning-guided approach provides a different measure of the Hubble tension.

\newpage

\tableofcontents 

\section{Introduction}
In this paper, we present a framework based on using a neural network to infer cosmological parameters from the Pantheon+ dataset
of \cite{Brout_2022} and a recent collection of BAO dataset presented in \cite{Bousis}. The neural network model we used is a surrogate model quantifying the luminosity distance $L$ vs redshift $z$ relationship, and its weight parameters are obtained through maximizing the degree of adherence to the following one-dimensional ODE
\be
\label{defining_eqn}
\frac{dL}{dz} - \frac{L}{1+z} - \frac{c(1+z)}{H(z; \vec{\Omega})} = 0,
\ee
where $H(z; \vec{\Omega})$ is the Hubble function parametrized by $\vec{\Omega}$ and $c$ is the speed of light. Eqn.~\eqref{defining_eqn} follows from the defining relation 
\be
\label{luminosity}
L = c(1+z)\int^z_0 d\tilde{z}
\frac{1}{H(\tilde{z}; \vec{\Omega})},
\ee
where $L$ is in units of Mpc, and its form is more convenient for us to infer unknown parameters $\vec{\Omega}$ of $H(z;\vec{\Omega})$ which depend on the underlying cosmological model assumed. In this work, we will consider two classes of deviations from the standard $\Lambda$CDM model described via the $w$CDM and $\Lambda_s$CDM models. 
The former refer to the standard $\Lambda$CDM model but with the equation of state parameter for dark energy $w$ not necessarily $-1$ (see e.g. \cite{wCDM} for a recent study).
In fitting it to the data, we take the free parameters of the $w$CDM model to be $\{H_0, \Omega_m, w \}$, 
where $H_0, \Omega_m$ are the Hubble constant and total matter density respectively.
For the $\Lambda_s$CDM model \cite{lambdaS}, here we take its free parameters to be $\{H_0, \Omega_m, z_t \}$ where $z_t$ is a transition redshift value from which the cosmological constant switches sign representing a toy model of vacua transition from anti-de Sitter to de Sitter spacetime at some point in the early universe.
These models are parametrically deformable to the standard $\Lambda$CDM model in the limits 
of $w \rightarrow -1$ for the $w$CDM model and $z_t \rightarrow \infty$ for the $\Lambda_s$CDM model. 

In standard regression techniques invoking the principle of maximum likelihood estimation, cosmological parameters are inferred through minimizing a $\chi^2$ likelihood of the form (see e.g. \cite{Trotta})
\be
\label{standard_reg}
-2 \log (\mathcal{L}) = \chi^2
 = \Delta \vec{D}^T C^{-1} \Delta \vec{D},
\ee
where $C$ is the covariance matrix expressing uncertainties and $\Delta D_k \equiv L_k - L_{model}(z_k)$ is the parameter residuals with $L_k$ being an observed value and $L_{model}(z_k)$ being the corresponding theoretical estimate computed with eqn.~\eqref{luminosity}. 

A fundamental difference between using \eqref{standard_reg} and a neural network-based approach is that the latter is structured around a surrogate model $\mathcal{M}(z)$ that represents the target variable as a function of the input variable, apart from an inference of the unknown parameters. Analogous to the pure numerical solution equipped with the best-fit parameters that minimizes  \eqref{standard_reg}, the final model $\mathcal{M}(z)$ is characterized by a set of network parameters that correspond to the minimum of a loss function that generalizes \eqref{standard_reg}. For a multilayer-perceptron model trained using just a mean-squared error loss term, model training then translates to solving \eqref{standard_reg} through a gradient-descent-based approach with $L_{model}$ being the neural network. For more complicated frameworks such as that of \emph{Physics-Informed Neural Networks} (PINN) \cite{pinn} where PDE constraints are simultaneously imposed, the loss function can be much more complex than \eqref{standard_reg}. 
In this work, we examine the use of $\mathcal{M}(z)$ as an independent data-driven model to infer probability distributions for the parameters from data, in a manner consistent with Bayesian principles. To do so, we need a framework that ideally yields $\mathcal{M}(z)$ together with its predictive uncertainty. It should also yield the posterior distribution for each unknown parameter of the cosmological model upon completion of model training.

Recently in \cite{tan_E,tan_S}, \emph{Evidential Physics-Informed Neural Networks} (or E-PINN for short) was proposed as a framework for PDE-based scientific modeling that encapsulates uncertainty quantification robustly. It realizes a hybrid implementation of the algorithms of \emph{Evidential Deep Learning} \cite{amini,sensoy} and those of PINN. In \cite{tan_S}, a principled approach was proposed for constructing priors for the unknown parameters and the learnable loss weight of the PDE residual term which is taken as the likelihood function for the unknown parameters. Gradient-descent based training then translates to the \emph{maximum a posteriori} learning of the distribution of the unknown parameters and weights of the surrogate model $\mathcal{M}(z)$. 
In this paper, we will use E-PINN as the machine learning framework for learning cosmological parameters from the Pantheon$+$ and BAO datasets. 

We examine the differences in the inferred cosmological parameters when E-PINN is trained on these datasets separately and examine how E-PINN differentiates among the alternative cosmological models with respect to each dataset and their synthesis. In the aspect of the model training algorithm, while still leveraging the basic framework of E-PINN as proposed in \cite{tan_S},
we also incorporate Gaussian Process regression \cite{GP} within the training algorithm in a few ways to refine the parameter inference process. Gaussian Process is used to guide the construction of the prior distributions for $\vec{\Omega}$. The predictive variance values provided by Gaussian Process regression are employed as proxy targets to supervise the learning of epistemic uncertainty in our model.
Although our primary motivation for incorporating Gaussian Processes into the framework stems from the relatively small size of the BAO dataset \cite{Bousis}, the methods we propose are readily transferable to other scientific modeling problems and extend the versatility of the E-PINN toolkit of \cite{tan_S}.

Previous to our work, there has been studies \cite{rover,Augusto,Qi,Wang,li,muk,abe} related to the use of neural network-based models for analyzing cosmological data and inference of parameters. In \cite{rover}, PINN was applied to Union 2.1 dataset and an uncertainty framework was proposed where the perceptron model's outputs were taken to be the (mean) luminosity distance and its associated uncertainty, with the loss function being the log-likelihood of the Gaussian with the outputs as its moments. For us, following the framework of Evidential Deep Learning, we assert a prior distribution (normal-inverse-gamma) for the mean and variances, integrating them out to obtain a t-distribution as the marginal likelihood. Our model's outputs then correspond to the learnable parameters of this higher-order distribution. In contrast to our work, in \cite{rover}, there was no methodology proposed to infer unknown cosmological parameters from the data. In \cite{Augusto},
the authors essentially used PINN (see eqn. 33 of \cite{Augusto}) and was focused primarily on whether PINN can be used to reproduce numerical solutions of PDE (in the context of cosmological models). \cite{Augusto} performed parameter inference but it was done using the standard regression method of minimizing the $\chi^2$. In \cite{Qi,Wang_2020}, no PINN-related formalism was invoked but a perceptron model trained on simulated data generated based on some chosen fiducial values of the cosmological parameters and synthetic noise added to the redshift. The statistical inference was done using the standard $\chi^2$ method as in \cite{Augusto}, rather than through a learned posterior distribution supported by a data-informed prior in our framework.

Our paper is organized as follows. We begin by presenting the theoretical formulation of the E-PINN model in Sec.~\ref{sec:model}, including how we invoked Gaussian Processes to enhance the original framework of \cite{tan_S}. This is followed by a discussion on methodology such as model training implementation details, metrics, etc. in Sec.~\ref{sec:method}. Our main results on the cosmological parameters are collected in Sec.~\ref{sec:results}. We end with a summary and some comments on the relevance of our work to the Hubble tension problem \cite{HTension} in Sec.~\ref{sec:conclusion}. Appendix~\ref{sec:AppA} contains some technical details related to the loss function, including
a detailed derivation for the hyperparameters of the prior for the PDE residual loss weight, while Appendix~\ref{sec:AppB} gathers various corner plots for the posterior distributions predicted by our models. Our study illustrates how a data-driven machine learning approach can be suitably adapted for cosmological parameter inference.

\section{Model formulation}
\label{sec:model}
In this Section, we introduce the main ideas and practical implementation of E-PINN, and explain how we extend the original algorithm of \cite{tan_S} by incorporating Gaussian Processes to construct the parameters' prior and supervise learning of the epistemic uncertainty. We refer the reader to \cite{tan_S} for a more technical exposition of E-PINN.

\subsection{Basic structure of the neural network and the loss function}

For our purpose (and for the general context of regression), we take the base neural network of E-PINN to be a multilayer perceptron $\mathcal{M}(z,\vec{w})$, where $z$ denotes its input,
and $\vec{w}$ its weight parameters. In our work here, $z$ is the redshift value associated with the measured luminosity distance of some astrophysical object/phenomenon. 
Our model $\mathcal{M}(z,\vec{w})$ has four output neurons $\{\alpha, \beta, \nu, \gamma\}$ with one of them ($\gamma$) pertaining to the mean of the target (luminosity distance) and the 
other three related ($\alpha, \beta, \nu$) to the predictive uncertainty as follows.
\be
\label{pred_uncertainty}
\sigma^2_p = \frac{\beta}{(\alpha - 1)}\left(
\frac{1}{\nu} + 1
\right).
\ee
The uncertainty $\sigma^2_p$ is the variance of a probability distribution that describes the statistical fluctuation of the observed target about its mean. This probability distribution is a t-distribution defined by the following density function. 
\be
\label{t_pde}
 P(\mathcal{D} | \mathcal{M}(\vec{w}) )
 =  
\frac{\Gamma\left( \alpha + \frac{1}{2} \right)}{\Gamma \left( \alpha  \right)  \sqrt{2\pi \beta  (1+\nu)/\nu  }}
\left(
1 + \frac{(L_{obs} - \gamma)^2}{2\beta  (1+\nu)/\nu}
\right)^{-(\alpha+\frac{1}{2})},
\ee
where $L_{obs}$ is the observed luminosity distance and $\gamma$ is the model prediction (to be interpreted as a mean value), with $\alpha, \beta, \nu$ being its other parameters which set the scale for its variance (eqn.~\eqref{pred_uncertainty}) and other moments. (In Appendix \ref{sec:AppA}, we provide more details on how this is related to Gaussian distributions.) In eqn.~\eqref{t_pde},  $P(\mathcal{D} | \mathcal{M}(\vec{w}) )$ denotes the probability of observing the data $\mathcal{D}$ given the model $\mathcal{M}(\vec{w})$. Here, and henceforth, we have suppressed the input $z$ for simplicity, and all variables on the RHS of eqn.~\eqref{t_pde} are to be understood as functions of $z$ which is the input variable of the model, with $\alpha,\beta,\nu,\gamma$ being the model's outputs. For them to be interpreted as what we have described thus far, we need a suitable loss function that is derived based on these principles. Here, we take the negative log-likelihood of eqn.~\eqref{t_pde} to be part of our complete loss function.
\be
\label{dataloss}
\mathcal{L}_{data} = - \log
\left[
P(\mathcal{D} | \mathcal{M}(\vec{w}) )
\right].
\ee
If we do not allude to any 
PDE constraints, i.e. asserting that eqn.~\eqref{defining_eqn} describes the data, then $\mathcal{L}_{data}$ would constitute our complete loss function, and model training typically means using a gradient-descent based method to find weights $\vec{w}$ such that they minimize $\mathcal{L}_{data}$. We call the set of points $(z_{obs},y_{obs})$ the training dataset, with model training translating to iterations in the gradient-descent based algorithm. 

Now, let us bring in the putative PDE description for the observed data. For this work, this would be eqn.~\ref{defining_eqn} with some ansatz for $H(z;\vec{\Omega})$ that is in turn derived from cosmological approximations in General Relativity, etc. With the parameters $\vec{\Omega}$ unknown, we add the following loss term
\be
\label{resi_exp}
\mathcal{L}_{pde} = -\log\left[
P(\mathcal{M}(\vec{w}) | \vec{\Omega} ) 
\right],\,\,\,
P(\mathcal{M}(\vec{w}) | \vec{\Omega} ) \sim
\exp{ \left[- \frac{1}{2\sigma^2_R}
\sum^{N_D}_{k=1}
\mathcal{R}^2_k \left(
\partial \gamma, \gamma, z_k, \vec{\Omega}
\right) \right]
},
\ee
where $k$ is a subscript denoting each data point, $N_D$ is the number of observations, and suppressing the data point index, we have defined 
$$
\mathcal{R} \left(
\partial \gamma, \gamma, z, \vec{\Omega}
\right) \equiv \frac{d\gamma}{dz} - \frac{\gamma}{1+z} - \frac{c(1+z)}{H(z; \vec{\Omega})},
$$
which refers to the RHS of eqn.~\ref{defining_eqn}, with $\gamma$ representing the mean luminosity distance. The parameter $\sigma^2_R$ is a learnable parameter that evolves during model training, and from  
eqn.~\eqref{resi_exp}, we can see that it acts as a weighting factor quantifying the magnitude of this loss term relative to other loss terms in the complete loss function. Eqn.~\eqref{resi_exp} guides the model towards adhering to the ODE \eqref{defining_eqn} when added to the complete loss function to be minimized. 
We have also used the symbol $P(\mathcal{M}(\vec{w}) | \vec{\Omega} )$ to denote the interpretation of this term as the 
likelihood of obtaining $\mathcal{M}(\vec{w})$ as a surrogate model assuming the parameters $\vec{\Omega}$.

In standard PINN, $\sigma^2_R$ is a free parameter and, to our knowledge, there is no principled approach towards determining its choice. For our E-PINN model, we lift $\sigma^2_R$ to be a learnable parameter that evolves as the model shifts towards a minimum in the loss landscape. We regularize this procedure by introducing another loss term that represents the negative logarithm of the prior density function for $\sigma^2_R$ as follows. 
\be
\label{prior_R_main}
\mathcal{L}_{\pi(\sigma^2_R)} 
= -\log \pi(\sigma^2_R; \alpha_r, \beta_r),\qquad
\pi(\sigma^2_R; \alpha_r, \beta_r) \equiv
\frac{\beta_r^{\alpha_r}}{\Gamma (\alpha_r)}
\sigma^{-2(\alpha_r + 1)}_R e^{-\frac{\beta_r}{\sigma^2_R}},
\ee
In Appendix \ref{sec:fix_sigmaR}, we furnish more detailed explanations of how \eqref{prior_R_main} is derived, in particular showing one can set $\alpha_r, \beta_r$ of \eqref{prior_R_main} such that these values align consistently with other aspects of our framework. 

Thus far, we have motivated the inclusion of three different loss terms
$\mathcal{L}_{data}, \, \mathcal{L}_{pde}, \,
\mathcal{L}_{\pi(\sigma^2_R)}$, with each of them interpretable as the negative logarithm of some density/likelihood function. If we follow principles of Bayesian statistics, we can further consider incorporating a prior density function $\pi(\vec{\Omega})$ for the unknown parameters. In Appendix \ref{sec:AppA_para_prior}, we show such a prior can be derived from considering the family of numerical solutions to the ODE and the model that is trained purely on data based on $\mathcal{L}_{data}$. 
Corresponding to $\pi(\vec{\Omega})$, we thus introduce another loss term 
$\mathcal{L}_{\pi(\Omega)} = -\log( \pi(\Omega) )$. 
Taking into account all the loss terms considered so far, we have the loss function
\bea
\label{loss_m}
\mathcal{L} &=& \mathcal{L}_{data} + 
\mathcal{L}_{pde} + \mathcal{L}_{\pi(\sigma^2_R)} + 
\mathcal{L}_{\pi(\Omega)}
\cr
&=& -\log\left[
P(\mathcal{D} | \mathcal{M}(\vec{w}) ) \,
P(\mathcal{M}(\vec{w}) | \vec{\Omega} )  \,\pi(\sigma^2_R; \alpha_r, \beta_r)\,
\pi ( \vec{\Omega} ) \right], 
\eea
with the product
$P(\mathcal{D} | \mathcal{M}(\vec{w}) ) \,
P(\mathcal{M}(\vec{w}) | \vec{\Omega} )$ being the joint likelihood function for $\vec{\Omega}$. Upon completion of model training, we can place confidence intervals on model's predictions using $\sigma^2_p$ of eqn.~\eqref{pred_uncertainty}. Since we infer the posterior distribution of $\vec{\Omega}$ at the end of model training, our framework thus appears as a \emph{maximum a posteriori} estimation of $\vec{\Omega}$, or more precisely\footnote{The posterior density implied by our loss function is not normalized, yet the normalization factor would involve $\vec{w}$ which is not taken into account during model training. For this reason, we consider our inference procedure a maximum likelihood estimation regularized by a prior density. } a MLE estimation that regularized by a prior $\pi (\vec{\Omega})$. 

From eqn.~\eqref{t_pde}, we can see that while the empirical data $L_{obs}$ correlates with the mean output $\gamma$ (through the factor $(L_{obs} - \gamma)^2$), there is no other external information supervising the three other uncertainty-related outputs $\alpha, \beta, \nu$. In our context, the cosmological datasets are already equipped with uncertainty estimates which we can potentially use to refine model training. In the following Section~\ref{sec:supervise_alea}, we discuss this issue in detail and introduce Gaussian Processes as a complementary tool to supervise the learning of the model's uncertainty. 
For model training, this further introduces a couple more loss terms to be added to \eqref{loss_m} for our complete loss function.

\subsection{Using Gaussian Processes to supervise uncertainties}
\label{sec:supervise_alea}

Although model training can proceed without additional information on data uncertainties, the datasets selected for our work here are already equipped with measurement uncertainties -- for the BAO data, these were computed in \cite{Bousis} from raw uncertainties of each sample as collected in Table 1 of \cite{Bousis}, whereas for Pantheon data, we used the diagonal elements of the covariance matrix presented in \cite{Scolnic_2022}. They correspond to what is known in the machine-learning terminology as `aleatoric uncertainties' which generally refer to uncertainties that are measurement or observation-related such as noise, etc. 
We add a simple mean-squared-loss term in the form 
\be
\label{alea_loss_m}
\mathcal{L}_{alea} = \mathbb{E} \left(
\frac{\beta}{\alpha - 1} - \sigma^2_a \right),
\ee
where $\sigma^2_a$ denotes the measurement's statistical variance for each point, and $\mathbb{E}$ denotes taking the average over all the training samples. In Appendix \ref{sec:AppA}, we explain more carefully why the factor of $\frac{\beta}{\alpha - 1}$ arises in \eqref{alea_loss_m}. 

Now there is another notion of uncertainty known as the `epistemic uncertainty' -- which pertains to the model itself instead of the data measurement process. In contrast to aleatoric uncertainty, this is a quantity that characterizes the degree of data sufficiency and model complexity. It can be supervised in model training if there is some independent knowledge of the model variance. Here, we use a Gaussian Process Regression model to furnish information on the epistemic uncertainty distribution. A Gaussian Process (GP) is essentially a distribution over functions.\cite{GP} Denoting the GP by $f(z)$, schematically
$$
f(z) \sim \mathcal{G} \mathcal{P} 
\left( m(z), k(z,z') \right),
$$
where $m(z)$ is the mean function and 
$k(z,z')$ is the covariance kernel function. Here we took $k(z,z') = \text{exp}\left( - (z-z')^2/2l^2 \right)$, a RBF function with a characteristic length scale $l$ that we determine by maximizing the log marginal likelihood using the implementation in scikit-learn \cite{scikit}. This length scale can be understood as a factor that controls the rate at which the correlation (as modeled by the kernel) decays with increasing separation. A relatively larger choice of $l$ implies a kernel that is smoother and more slowly varying. Conditioned on an observed set of data 
$\{ z_i, y_i \}^{N_D}_{i=1}$, the posterior distribution for $f$ evaluated at some arbitrary redshift $\tilde{z}$ is a Gaussian distribution 
$\mathcal{N}(\tilde{\mu}, \tilde{\sigma}^2_e )$
with moments 
\be
\label{aux_Gaussian}
\tilde{\mu} = k(\tilde{z},z) [k(z,z) + \sigma^2_a \, \mathbb{I} ]^{-1} y, \,\,\,
\tilde{\sigma}^2_e = k(\tilde{z}, \tilde{z}) - 
k(\tilde{z},z) [k(z,z) + \sigma^2_a \, \mathbb{I} ]^{-1} k(z, \tilde{z}),
\ee
where $\sigma^2_a$ is the aleatoric uncertainty and $\sigma^2_e$ is used to supervise the learning of the epistemic uncertainty. 
To see why this is a natural choice, we recall that our framework assumes an auxiliary Gaussian target (the luminosity distance in our context) with normal-inverse-gamma distribution being the prior for its mean and variance, and the epistemic uncertainty is the expectation value of the auxiliary Gaussian's variance. Thus, the GP variance  $\tilde{\sigma}^2_e$ in \eqref{aux_Gaussian} is a natural candidate for supervising the learning of the epistemic uncertainty. Like aleatoric uncertainty in \eqref{alea_loss_m}, we introduce an additional mean-squared loss term of the form 
\be
\label{epi_loss_m}
\mathcal{L}_{epi} = \mathbb{E} \left(
\frac{\beta}{\nu (\alpha - 1)} - \sigma^2_e
\right),
\ee
where $\sigma^2_e$ is the GP variance at each training datapoint, and we are averaging over the training dataset. 
In Appendix \ref{sec:AppA}, we explain more carefully how the factor of $\frac{\beta}{\nu(\alpha - 1)}$ arises in \eqref{epi_loss_m}. 
Adding both uncertainty-related losses to eqn.~\eqref{loss_m}, our complete loss function is thus
\bea
\label{complete_loss_m}
\mathcal{L} &=& \mathcal{L}_{data} + 
\mathcal{L}_{pde} + \mathcal{L}_{\pi(\sigma^2_R)} + 
\mathcal{L}_{\pi(\Omega)} + \lambda_a \mathcal{L}_{alea} 
+ \lambda_e \mathcal{L}_{epi}
\cr
&=& -\log\left[
P(\mathcal{D} | \mathcal{M}(\vec{w}) ) \,
P(\mathcal{M}(\vec{w}) | \vec{\Omega} )  \,\pi(\sigma^2_R; \alpha_r, \beta_r)\,
\pi ( \vec{\Omega} ) \right]+ \lambda_a \mathcal{L}_{alea} 
+ \lambda_e \mathcal{L}_{epi}.
\eea
The addition of the loss terms \eqref{alea_loss_m} and \eqref{epi_loss_m} guides the learning of the uncertainty-related model outputs $\alpha, \beta, \nu$ to complement how the observed data supervises the learning of the mean target variable $\gamma$. We weighted each loss term with tunable coefficients $\lambda_e, \lambda_a$ that can be adjusted as hyperparameter to yield a good error calibration at the end of model training.

\subsection{A summary list of model implementation }
\label{sec:guide}
For clarity, in the following, we provide a brief overview of the model implementation process. Our framework is structured around a two-phase training algorithm where in the first phase, the neural network is trained purely on the empirical dataset without alluding to the ODE in eqn.~\ref{defining_eqn}. Apart from being used to derive one of the loss term ($\mathcal{L}_{\pi(\sigma^2_R)}$), this initial model also furnishes the general initial conditions (e.g. initial values for some of the learnable parameters, etc.) for model training in the second phase which uses the full loss function of \eqref{complete_loss_m}.

\begin{enumerate}[label=(\Roman*)]
\item
The loss function in this training phase is consists of three loss terms: the data loss term \eqref{dataloss}, the aleatoric \eqref{alea_loss_m} and epistemic \eqref{epi_loss_m} loss terms. 
\be
\mathcal{L}_{\text{1st phase}} = 
-\log\left[
P(\mathcal{D} | \mathcal{M}(\vec{w}) ) \right] + \lambda_a \mathcal{L}_{alea} 
+ \lambda_e \mathcal{L}_{epi}.
\ee
\item
Independently, a GP regression model is fitted to data so as to gain epistemic uncertainty information for supervising $\mathcal{L}_{epi}$, and for constructing $\pi(\vec{\Omega})$ (see Appendix~\ref{sec:AppA_para_prior} for full details)
\item
Upon convergence of the purely data-fitted model, 
we can now determine the prior $\pi(\sigma^2_R; \alpha_r, \beta_r)$ 
following the steps described in Appendix \ref{sec:fix_sigmaR}. 
(This essentially involves solving
for $\alpha_r, \beta_r$ using \eqref{sigmaR}, \eqref{sigmaR_mode}.)  
\item
We then proceed with the second phase of model training. This phase of training refines the purely data-fitted model such that it conforms to the presumed PDE description. Apart from the model's weights, $\{\vec{\Omega}, \sigma^2_R\}$ are also learnable parameters. 
In this final phase, the model is trained using the full loss function 
\be
\label{full_loss}
\mathcal{L} = -\log\left[
P(\mathcal{D} | \mathcal{M}(\vec{w}) ) \,
P(\mathcal{M}(\vec{w}) | \vec{\Omega} )  \, 
\pi (\sigma^2_R; \alpha_r, \beta_r) \,
\pi ( \vec{\Omega} ) \right] + \lambda_a \mathcal{L}_{alea} 
+ \lambda_e \mathcal{L}_{epi},
\ee
with each of the six individual loss terms defined in eqns.~\eqref{dataloss}, \eqref{resi_exp}, \eqref{prior_R_main}, \eqref{alea_loss_m}, \eqref{epi_loss_m} and \eqref{prior_formula_1}. 
\item
Upon completion of training, the model predictions are expressed by the target variable $\gamma$ while confidence bands can be constructed from $\alpha, \beta, \nu$. We also infer the PDE parameters $\vec{\Omega}$ with its uncertainty as defined by the median and credible intervals of the posterior distribution. 
\end{enumerate}

In \cite{tan_S}, this framework has been validated and compared against two other more popularly known models (Bayesian Physics-Informed Neural Networks
and Deep Ensemble) for its uncertainty quantification and accuracy in recovering unknown PDE parameters. The controlled case studies used in \cite{tan_S} were nonlinear second-order differential equation in 1D (Poisson equation with Gaussian source) and 2D (Fisher-KPP equation). As shown in \cite{tan_S}, the most significant superiority of E-PINN over these two standard uncertainty quantification frameworks is that it yields uncertainty estimates that are calibrated much more consistently (i.e. the confidence intervals deduced are much more consistent with actual model errors). In our context of inferring cosmological parameters, E-PINN thus appears to be well-suited, since having a robust uncertainty quantification framework that can generate reliable posterior distributions is particularly critical for our purpose.

\section{Methodology}
\label{sec:method}
\subsection{On the datasets and some limitations}

The BAO dataset collected in Table 1 of \cite{Bousis} consists of 32 measurements. It is a list of transverse BAO measurements of the comoving angular diameter distance $D_M/r_d$, where $r_d$ is the sound horizon scale at the end of the baryonic drag epoch. 
These samples includes recent data such as those made by DESI \cite{B72,B77}, the Sloan Digital Sky Survey (SDSS)
\cite{B73,B74,B76,B78,B79,B80,B82,B83,B84,B85,B86,B87,B88,B89} and the Dark Energy Survey (DES) \cite{B75,B81}.
As described in \cite{Bousis}, the samples involved anisotropic BAO analyses which incorporate the full 3D galaxies' distributions, often based on some fiducial cosmological model to convert observed angles and redshifts into physical distances. In \cite{Bousis}, the sound horizon $r_d$ was taken to be 147.18 Mpc following Planck18 report \cite{Planck18}, and here we adopted the same value for $r_d$ in when translating values of $D_M/r_d$ in Table 1 of \cite{Bousis} to $D_L$. This is a limitation of our work which, in principle, can be overcome by deriving expressions for $r_d$ for each cosmological models (equipped with unknown, learnable parameters) and then replacing numerical luminosity distance targets with $r_d(\vec{\Omega}) \times (1+z) N_{data}$ where $N_{data}$ is the numerical $D_M/r_d$ value in Table 1 of \cite{Bousis}. In practice, this would complicate the gradient-descent based model training because $r_d(\vec{\Omega})$ can only be expressed through a numerical integral and not an explicit function of $\vec{\Omega}$. An ideal approach would be adopt a model-independent value for $r_d$ if possible. Interestingly, we note that lowering the sound horizon to 140 Mpc recently proposed by Liu et al. in \cite{liu_sound}) to be a model-independent result would naively yield the BAO dataset to be visually compatible with that of Pantheon$+$ data on the $(D_L, z)$ plane. 

In this work we have chosen to use the collection of BAO data of \cite{Bousis} for definiteness to enable comparison of the posterior distributions and best-fit values to the standard analysis which was performed in \cite{Bousis}. The data points presented in Table 1 of \cite{Bousis} are however characterized by correlations\footnote{We are grateful to an anonymous referee for advising us to emphasize this caveat.}, since several entries are obtained from surveys that partially overlap in sky coverage and source populations (e.g., galaxies and quasars). For example, the pair of points labeled $N=1,2$ in \cite{Bousis} with $(z,D_M/r_d)$ values being (0.32, 8.54), (0.32, 8.76) both originated from SDSS-III data. A limitation of our model is that it does not enable any external knowledge of correlations among points to be used directly in the learning of weights. The fundamental reason is that the t-distribution \eqref{t_distribution} underpinning the data loss term is derived from marginalizing over the means and variances of products of univariate Gaussian distributions (rather than multivariate ones) . On the other hand, we note that there are also pairs of points in Table 1 of \cite{Bousis} sharing the same redshifts but originating from different surveys, e.g. $N=5 \,(\text{BOSS}), 6\,(\text{DESI})$, 
$N=18\,(\text{DES}), 19\,(\text{eBOSS})$, etc. 

The Pantheon$+$ dataset \cite{Brout_2022} provides Type Ia supernovae (SNe Ia) luminosity distance and distance moduli measurements for redshifts in the range $z \in [0.001, 2.3]$, calibrated by the second rung of the distance ladder using Cepheids with the absolute magnitude being $M_B = -19.25 \pm 0.01$. The samples consists of 1701  light curves of 1550 spectroscopically confirmed SNe Ia. The data together with the uncertainties can be found at their \href{https://github.com/PantheonPlusSH0ES}{GitHub website}. 
A limitation of our usage of this dataset is that we only used the diagonal elements of the covariance matrix for supervising the aleatoric uncertainty. Similar to the case for the BAO dataset, any external knowledge of correlations among data points cannot be used in our formalism. There are no output variables of our neural network model that can interpreted as correlations (or their functions) between different data points. By construction, the four outputs of our model are related to the mean and uncertainty of the luminosity distance at each individual data point. Any generalization of our model to enable the full covariance matrix to be used would have to admit a much larger output layer with targets that are related to the correlations.

\begin{figure}[h!]
	\begin{center}		
\includegraphics[width=0.9\textwidth]{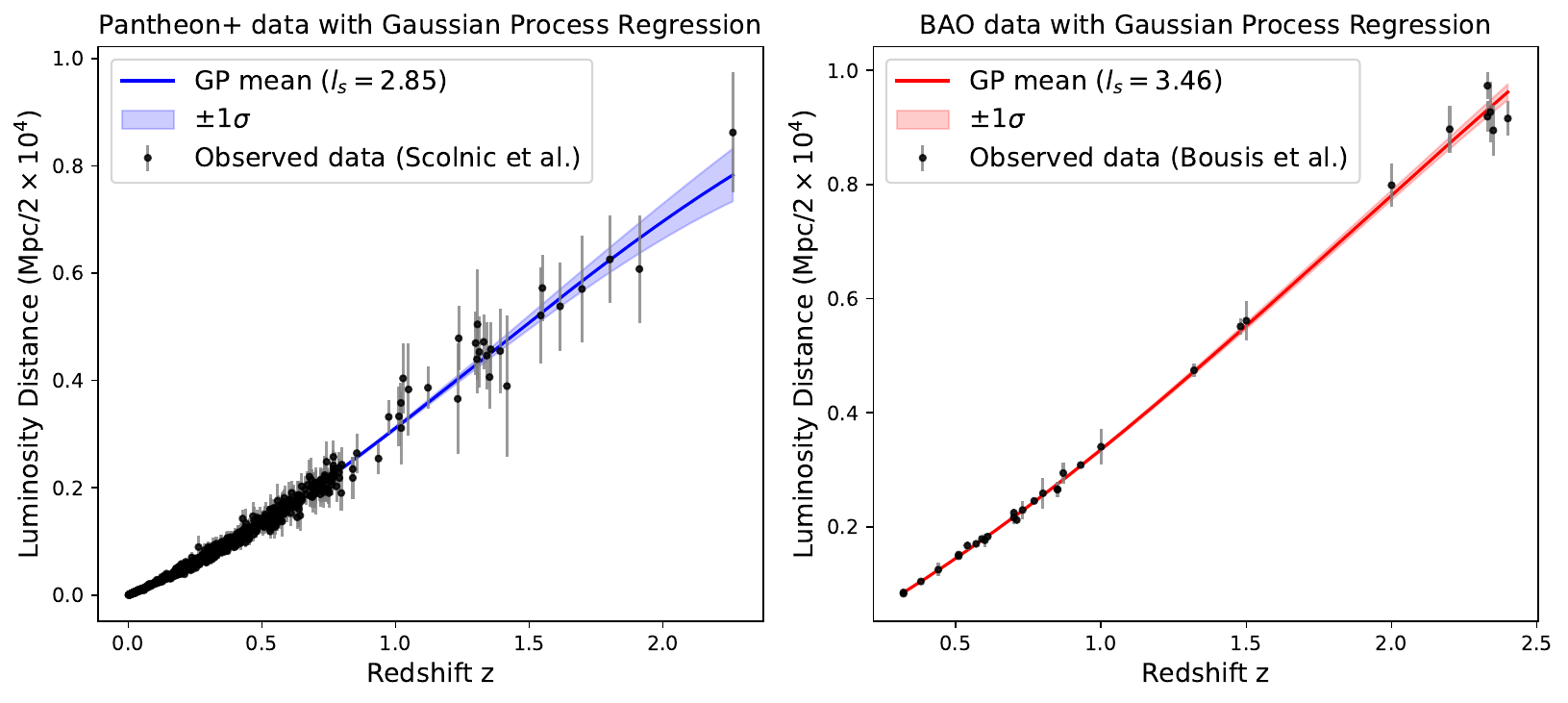}
	\end{center}
	\caption{\small Diagrams showing both the Pantheon$+$ and BAO datasets together with the fitted Gaussian Process Regression curves. The $1 \sigma $ confidence bands were used to supervise epistemic uncertainty whereas the empirical error bars were used to guide learning of aleatoric uncertainty in our model. }
	\label{fig:datasets}
\end{figure}

Each cosmological model is associated with a different Hubble function. In our work here, we set the curvature density term to be zero for simplicity, leaving generalizations that treat it as a learnable parameter for future work. The $w$CDM and $\Lambda_s$CDM models are defined as follows. 
\bea
\frac{H_{w\text{CDM}}(z)}{H_0}
&=&  \left( \Omega_{m}(1+z)^3
+ (1-\Omega_m)(1+z)^{3(1+w)} \right)^{1/2},\\
\frac{H_{\Lambda_s\text{CDM}}(z)}{H_0},
&=&
\left( \Omega_{m}(1+z)^3
+ (1-\Omega_{m}) \text{sgn}(z_t - z) \right)^{1/2},
\eea
where we have assumed the Planck-measured radiation density parameter $\Omega_{r} \sim 9.26 \times 10^{-5} \sim 0$ for simplicity. We take the free parameters of the $w$CDM model to be $\{H_0, \Omega_m, w \}$ where $w$ is the dark energy equation of state parameter. For $\Lambda_s$CDM model, its free parameters are taken to be $\{H_0, \Omega_m, z_t \}$ where $z_t$ is a transition redshift value from which the cosmological constant switches sign representing a toy model of vacua transition from anti-de Sitter to de Sitter spacetime at some point in the early universe. For computational convenience, here we use a hyperbolic tangent function as a smooth representation of the signum function. 
These models are parametrically deformable to the standard $\Lambda$CDM model in the limits 
of $w \rightarrow -1$ for the $w$CDM model and $z_t \rightarrow \infty$ for the $\Lambda_s$CDM model.

\subsection{Model training setup and implementation details}
In the following, we furnish some details of the model training, organizing them in terms of the dataset that was used. The base surrogate model $\mathcal{M}(z)$ was implemented as a fully-connected perceptron with two hidden layers of 32 neurons each and an output layer of 4 neurons corresponding to the 
$\{\alpha,\beta,\nu,\gamma\}$ variables of \eqref{t_distribution}. 
For each training dataset, the same initial model $\mathcal{M}_0$ was used for training the different cosmological model-based neural networks. The finite parameter domains were chosen to be 
$\Omega_m \in (0.10, 0.55), h_0 \in (0.50, 0.90), w \in (-2.0, -0.01), z_t \in (1.5, 3.5)$. We implemented Gaussian Process (GP) regression with a radial-basis function kernel via the scikit-learn library \cite{scikit}, with the optimized kernel's characteristic length-scales being 2.85 for the Pantheon$+$ data, 3.46 for the BAO data and 2.89 for the combined dataset.

For models trained on the Pantheon$+$ and the combined Pantheon$+$BAO datasets, 
in the initial training phase, we used a learning rate of $5\times 10^{-6}$
for the first $5 \times 10^4$ epochs and $10^{-6}$ for the subsequent ones with the total number of epochs being $10^6$. 
The data uncertainty hyperparameters were taken to be $\lambda_e  = \lambda_a = 1$.
For the second phase, the learning rate was $5\times 10^{-6}$ for the first $1.2\times 10^6$ epochs followed by $1\times 10^{-6}$ for another $1\times 10^6$ epochs. 
On the other hand, for the smaller BAO dataset, convergence was attained for various cosmological models in $3\times 10^5$ epochs with a learning rate of $2 \times 10^{-5}$ for the first $2\times 10^5$ epochs and followed by $2\times 10^{-6}$ for the remaining ones. The data uncertainty hyperparameters were taken to be $\lambda_e  = \lambda_a = 10^8$. 
Final relative tolerance was of the order $\sim 10^{-7}$ for Pantheon$+$ data-based models and higher at $\sim 10^{-5}$ for BAO data-based ones.

Table \ref{tab:prior} collects the parameters' prior densities for each model. These parameters were determined from the empirical distribution that measures the likelihood of each parametrized family of numerical solutions of the PDE using its deviations from the corresponding purely data-fitted model.

\begin{table}[h]
\centering
{\small
\begin{tabular}
{p{3cm}|p{3.5cm}|p{3.5cm}|p{3.5cm}}
\hline
\hline
$\,$ & $\Lambda$CDM & $\Lambda_s$CDM & $w$CDM\\
\hline
Pantheon$+$ data & $\Omega_m = 0.357 \pm 0.164$, $h_0 = 0.729 \pm 0.119$ 
& 
$\Omega_m  = 0.357 \pm 0.164$,
$h_0 = 0.729 \pm 0.119$,
$z_t = 2.520 \pm 0.755$
&  $\Omega_m = 0.376 \pm 0.164$, $h_0 = 0.769 \pm 0.129$, $w = -1.553 \pm 0.725$ \\ 
\hline 
BAO data & $\Omega_m = 0.357 \pm 0.160$, $h_0 = 0.671 \pm 0.127$
& 
$\Omega_m = 0.366 \pm 0.159$, $h_0 = 0.663 \pm 0.128$, $z_t = 2.643 \pm 0.752$
& 
$\Omega_m = 0.339 \pm 0.161$,
$h_0 = 0.720 \pm 0.137$,
$w = -1.472 \pm 0.714$
\\
\hline
Combined Pantheon$+$BAO & 
$\Omega_m = 0.238 \pm 0.163$,
$h_0 = 0.737 \pm 0.121$
&  
$\Omega_m = 0.247 \pm 0.163$,
$h_0 = 0.729 \pm 0.121$,
$z_t = 2.684 \pm 0.754$
& 
$\Omega_m = 0.256 \pm 0.163$,
$h_0 = 0.794 \pm 0.132$,
$w = -1.553 \pm 0.717$
\\
\hline
\end{tabular}}
\caption{{\small Prior density function for each parameter was taken to be univariate Gaussians of which means and standard deviations are tabulated here for all three models trained on each dataset. The means and variances are the modes and variances of $f(\vec{\Omega})$ so that the Gaussian priors are representative of the highest density regions of $f(\vec{\Omega})$.}}
\label{tab:prior}
\end{table}

\subsection{On empirical coverage probability and log model evidence}

Upon completion of model training, we assess the uncertainty quantification through computing the empirical coverage probability (ECP). The ECP at level $1-\alpha$
is the proportion of observed target values that fall within the corresponding t-distribution–based confidence band of \eqref{t_distribution}. To assess the degree of calibration, one can compare the ECP values to their nominal target level ($1-\alpha$) (nominal coverage probabilities). On the ECP vs NCP plane, a robust uncertainty quantification would yield a curve that is close to the straight line joining the origin to (1,1). A representative index would be 
the mean of the absolute discrepancy between the ECP and NCP. For each model, we computed this mean calibration error (MCE) (see also \cite{jungo}) and examined plots of ECP vs NCP, finding that all MCE are very small $\lesssim 0.05$, with models trained on Pantheon data better-calibrated with an MCE that is 0.1 smaller than those trained on BAO data. Most crucially, none of the 9 models had a ECP curve that is dominantly above or below the ideal line which would have indicated a systematic bias. 

\begin{figure}[h!]
	\centering
    \begin{subfigure}[b]{0.4\textwidth}
\includegraphics[width = 0.8\textwidth]{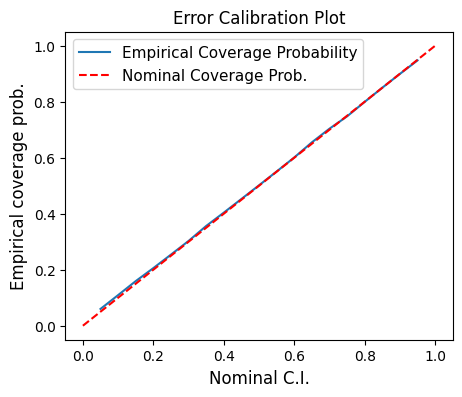}
	\caption{\small $\Lambda_s$CDM model (Pantheon data)  }
\end{subfigure}
  \begin{subfigure}[b]{0.4\textwidth}
\includegraphics[width=0.8\textwidth]{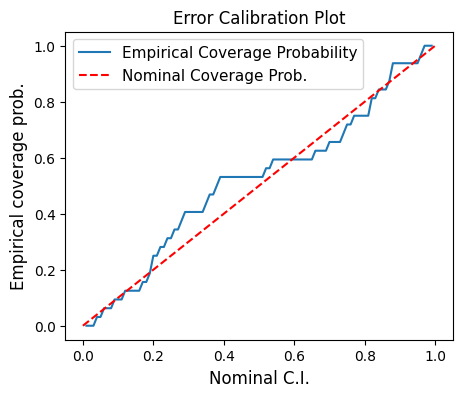}
	\caption{\small $w$CDM model (BAO data) }
    \end{subfigure}
\caption{\small Plots of empirical coverage probabilities vs their nominal values for a couple of models. The perfectly calibrated uncertainty-aware model would exhibit a straight line joining the origin to $(1,1)$. Models trained on BAO data exhibited less ideal ECP plots compared to those trained on Pantheon data, most likely attributable to the much smaller size of the dataset.  }
    \label{fig:ECP_diag}
\end{figure}

\begin{figure}[h!]
	\centering
    \begin{subfigure}[b]{0.7\textwidth}
\includegraphics[width = \textwidth]{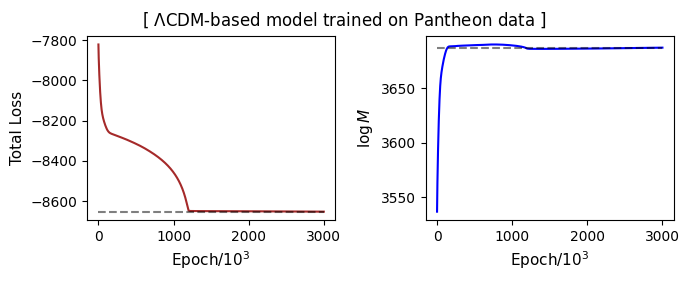}
\end{subfigure}
  \begin{subfigure}[b]{0.7\textwidth}
\includegraphics[width=\textwidth]{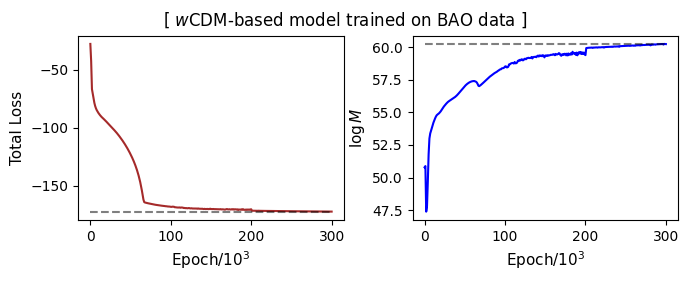}
    \end{subfigure}
\caption{\small Evolution of loss function and log $M$ for a couple of models (\emph{Top}: $\Lambda$CDM trained on Pantheon data; \emph{Bottom}: $w$CDM model trained on BAO data). All models have been checked to display convergence with a relative tolerance $<10^{-4}$ in both the loss and log $M$ term. }
    \label{fig:log_Bayes}
\end{figure}

The loss function of our model is, up to a normalization factor, the posterior distribution. The completion of model training yields quantities are directly related the log model evidence that can be further used to discriminate between models. Integrating out the parameters $\vec{\Omega}$, the model likelihood $M$ and its logarithm are
\bea
M &=& 
\int d\Omega\,\,\,
P(\mathcal{D} | \mathcal{M}(\vec{w}) ) \,
P(\mathcal{M}(\vec{w}) | \vec{\Omega} )  \,
\pi ( \vec{\Omega} ), \cr
\log M &=& \log P(\mathcal{D} | \mathcal{M}(\vec{w})) + \log 
\left(
\int d\Omega\,\,\,
P(\mathcal{M}(\vec{w}) | \vec{\Omega} )  \,
\pi ( \vec{\Omega} )
\right),
\eea
where $\vec{w}$ are the final model weights and biases. The log model evidence ($\log M$) is often used as a measure to quantify and compare support between competing statistical models from data \cite{Trotta,morey}.
In Table~\ref{tab:results_cosmo}, we display the log model evidence for each model as a comparison index among models trained on the same dataset. In Fig.~\ref{fig:log_Bayes}, we show the evolution of log $M$ for a couple of models together with the associated loss function. All models have been checked to display convergence with a relative tolerance $<10^{-4}$ in both the loss and log $M$ term.

\section{Results}
\label{sec:results}
We collect the inferred parameters together with their uncertainties in Table \ref{tab:results_cosmo} below. 
Generally, we found that for each class of cosmological models, the neural networks trained separately on Pantheon and BAO data exhibited a systematic difference evident in the residuals in their luminosity-redshift curves (see Fig.~\ref{fig:resi_1D}) and the inferred posterior distributions of the parameters (see Fig.~\ref{fig:joint_marginals} ). When trained on the combined dataset, all models yielded higher $h_0$ and lower $\Omega_m$ compared to when being trained only on Pantheon$+$ dataset, with $\Lambda_s$CDM being associated with a clearly lower log model evidence compared to $\Lambda$CDM and $w$CDM models.  

\renewcommand{\arraystretch}{1.3} 
\begin{table}[h]
\centering
\caption{\small Table of inferred parameters (posterior medians with 0.68 C.I.) and logarithm of model evidence ($\log M$). For each dataset (Combined, Pantheon$+$, BAO), the cosmological model with the highest $\log M$ is shaded in gray.} 
\label{tab:results_cosmo}
{\small
\begin{tabular}{
  >{\centering\arraybackslash}m{1.5cm}|
  >{\centering\arraybackslash}m{2.0cm}|
  >{\centering\arraybackslash}m{2.2cm}|
  >{\centering\arraybackslash}m{2.2cm}|
  >{\centering\arraybackslash}m{2.2cm}|
  >{\centering\arraybackslash}m{2.0cm}}
\hline
\hline
Model & Dataset & $h_0$ & $\Omega_m$ & $w \,(w\text{CDM})$, $z_t\,(\Lambda_s\text{CDM})$ 
& $\log M$\\
\hline\hline
$\,$ & Pantheon$+$ & 
$0.729^{+0.033}_{-0.024}$
& $0.357^{+0.101}_{-0.092}$
& $\,$
&3687
\\
$\Lambda$CDM & BAO & 
$0.680^{+0.090}_{-0.082}$
& $0.357^{+0.110}_{-0.110}$
& $\,$
& 59.8
\\
$\,$ &\cellcolor{gray!20}  Combined & \cellcolor{gray!20} 
$0.745^{+0.033}_{-0.041}$
& \cellcolor{gray!20}  $0.320^{+0.092}_{-0.092}$
& $\,$
& 3717
\\
\hline
$\,$ & Pantheon$+$ & 
$0.745^{+0.033}_{-0.024}$
& $0.385^{+0.073}_{-0.083}$
& $-1.431^{+0.406}_{-0.366}$
& 3685
\\
$w$CDM & BAO & 
$0.712^{+0.090}_{-0.098}$
& $0.348^{+0.110}_{-0.110}$
& $-1.310^{+0.528}_{-0.447}$
& 60.2
\\
$\,$ & Combined & 
$0.769^{+0.057}_{-0.057}$
& $0.293^{+0.110}_{-0.092}$
& $-1.350^{+0.447}_{-0.406}$
& 3712
\\
\hline
$\,$ & \cellcolor{gray!20} Pantheon$+$ & \cellcolor{gray!20} 
$0.729^{+0.033}_{-0.024}$
& \cellcolor{gray!20}  $0.357^{+0.101}_{-0.083}$
& \cellcolor{gray!20}  $2.520^{+0.571}_{-0.571}$
& 3688
\\
$\Lambda_s$CDM & \cellcolor{gray!20} BAO & 
\cellcolor{gray!20}  $0.671^{+0.090}_{-0.073}$
& \cellcolor{gray!20}  $0.366^{+0.110}_{-0.110}$
& \cellcolor{gray!20}  $2.602^{+0.531}_{-0.612}$
& 61.0
\\
$\,$ & Combined & 
$0.737^{+0.041}_{-0.033}$
& $0.274^{+0.101}_{-0.083}$
& $2.602^{+0.571}_{-0.571}$
& 3650\\
\hline
\end{tabular}}
\end{table}

\subsection{On tensions between models trained separately on Pantheon$+$ and BAO data}

We examined the difference in the joint marginal distributions of $h_0$ and $\Omega_m$ for the three models, each trained separately on the Pantheon$+$ and BAO data. Fig.~\ref{fig:joint_marginals} shows the 68\% and 95\% contours for each model. The $\Lambda$CDM and $\Lambda_s$CDM models yielded similar distributions 
with the Jensen-Shannon divergence \cite{JS} between the Pantheon and BAO data-based distributions being 2.495 and 2.592 respectively,
while that of $w$CDM model was characterized by the least Jensen-Shannon divergence of 2.342. The posterior distributions for the cosmological parameters in various models trained purely on Pantheon$+$ data were found to be largely contained within the $2\sigma$ contours of their counterparts trained on BAO data (Fig. \ref{fig:joint_marginals}). This is in stark contrast to Fig.~4 of \cite{Bousis} where the posterior distributions did not overlap at $3\sigma$.

\begin{figure}[h!]
    \centering
    \begin{subfigure}[b]{0.45\textwidth} \includegraphics[width=\textwidth]{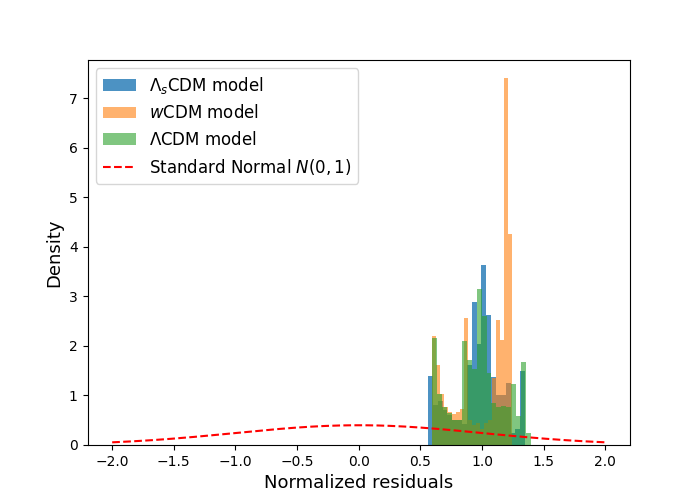}
        \caption{Normalized Residuals}
        \label{fig:norm_resi}
    \end{subfigure}
    \begin{subfigure}[b]{0.45\textwidth}
\includegraphics[width=\textwidth]{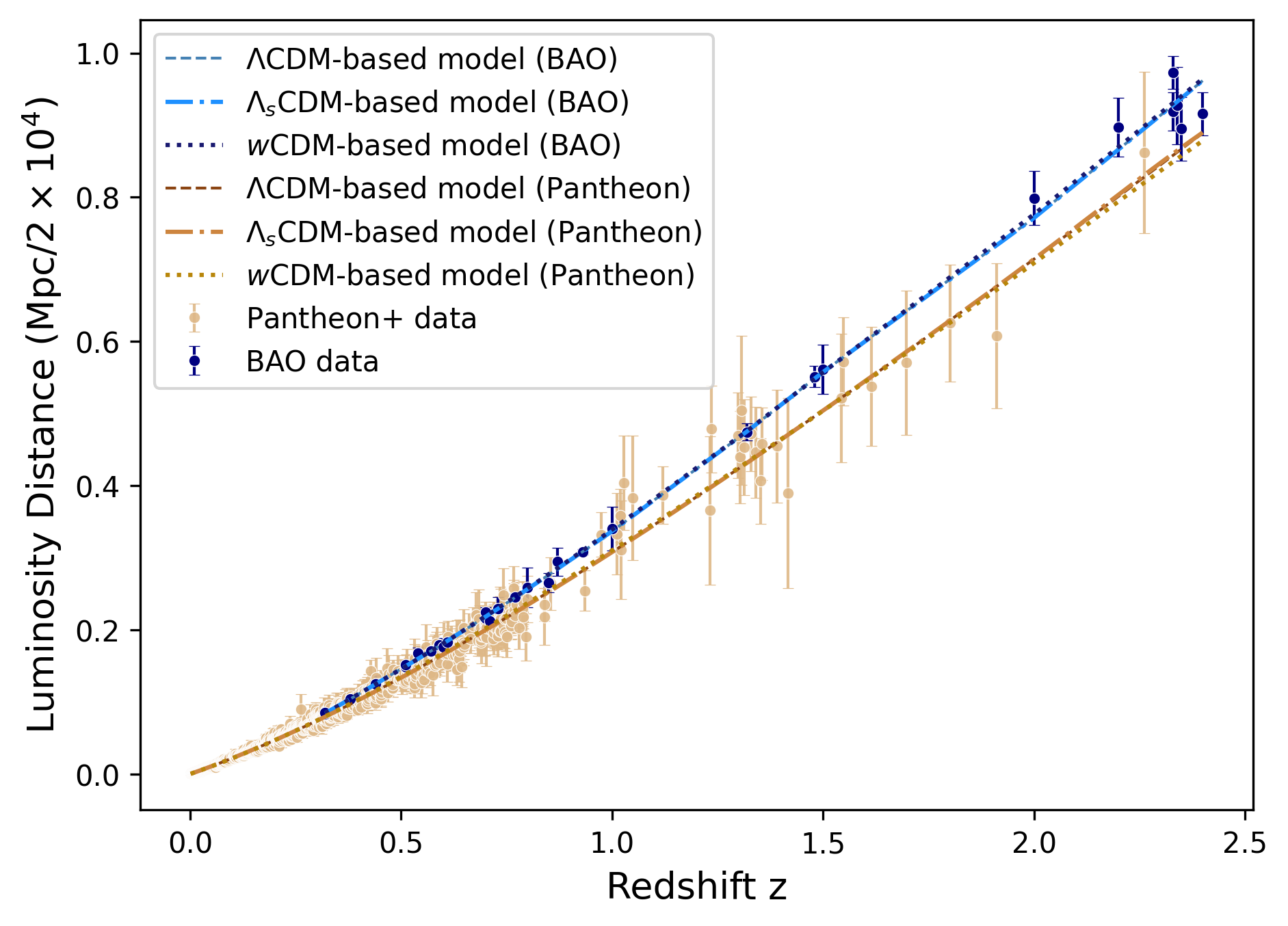}
        \caption{Model Prediction Curves}
        \label{fig:1D_s}
    \end{subfigure}
    \caption{\small Left diagram shows residuals between BAO data and Pantheon$+$ data-trained models, normalized by the combined model uncertainties, highlighting systematic differences induced by dataset choice. Right diagram collects all model predictions. }
    \label{fig:resi_1D}
\end{figure}

\begin{figure}[h!]
    \centering
    \begin{subfigure}[b]{0.30\textwidth}
        \includegraphics[width=\textwidth]{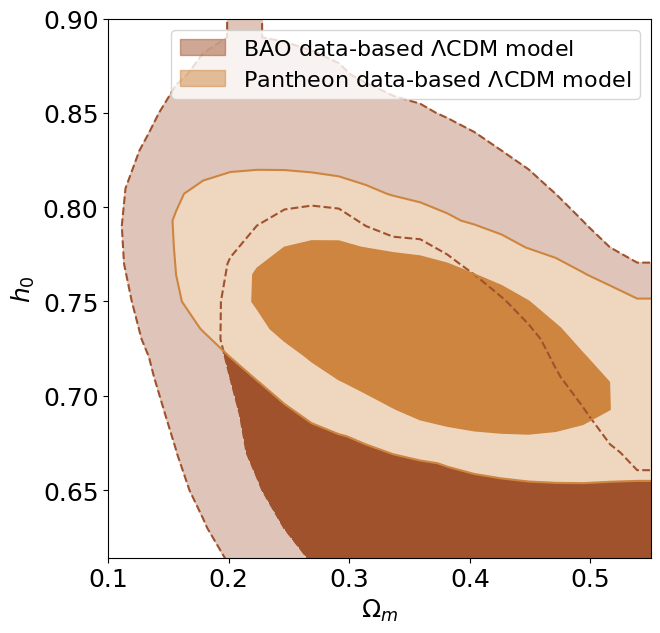}
        \caption{$\Lambda$CDM model}
        \label{fig:lambda_mar}
    \end{subfigure}
    \begin{subfigure}[b]{0.30\textwidth}
        \includegraphics[width=\textwidth]{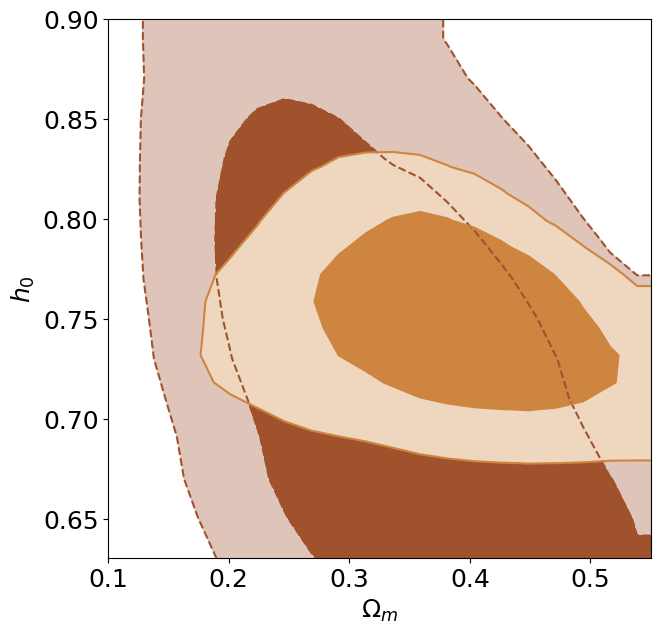}
        \caption{$w$CDM model}
        \label{fig:wcdm_mar}
    \end{subfigure}
    \begin{subfigure}[b]{0.30
    \textwidth}
        \includegraphics[width=\textwidth]{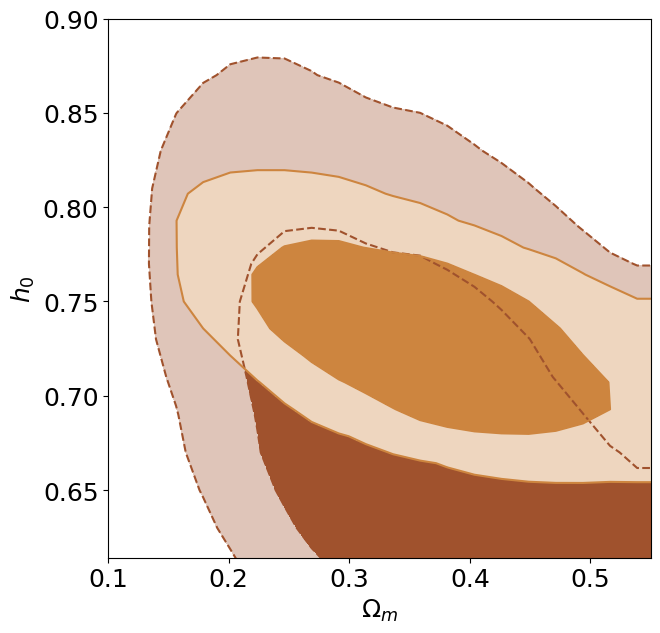}
        \caption{$\Lambda_s$CDM model}
        \label{fig:scdm_mar}
    \end{subfigure}
    \caption{\small Joint marginal distributions of \(h_0\) and \(\Omega_m\) for the three models, each trained separately on the Pantheon$+$ (light brown) and BAO data (dark brown). The 68\% and 95\% credible contours are shown for each panel. For the $w$CDM and $\Lambda_s$CDM models, the distributions shown were obtained after marginalizing over $w$ and $\Lambda_s$ parameters respectively. }
    \label{fig:joint_marginals}
\end{figure}

For each cosmological model, we consider the differences in the predicted luminosity-distance curves resulting from the model being trained purely on either BAO or Pantheon$+$ datasets. 
The normalized residuals between the predictions of the BAO-trained and Pantheon-trained models are shown in Fig.~\ref{fig:resi_1D}. 
We found that these residuals exhibited strong deviations from the  
$\mathcal{N}(0,1)$ distribution ($p \approx 0$) associated with statistical noise. This indicates the presence of dataset-dependent systematic effects, whereby each dataset favors a different best-fit model.

\subsection{On models trained on the combined Pantheon$+$ and BAO data}

When trained on the combined dataset, 
all three models yielded similar prediction curves as depicted in the Fig.~\ref{fig:pb_curves} below. 
The $\Lambda$CDM and $w$CDM models showed the highest log Bayes factor, and all three models yielded posterior medians of $h_0, \Omega_m$ that agree within one standard deviation. The posterior medians for $h_0$ for all models were all larger than $0.73$, with a standard deviation falling within $(0.03, 0.06)$. Each model yielded lower values of $\Omega_m$ and higher values of $h_0$ than when trained on the individual datasets separately.

\begin{figure}[h!]
	\centering
    \begin{subfigure}[b]{0.49\textwidth}
\includegraphics[width = \textwidth]{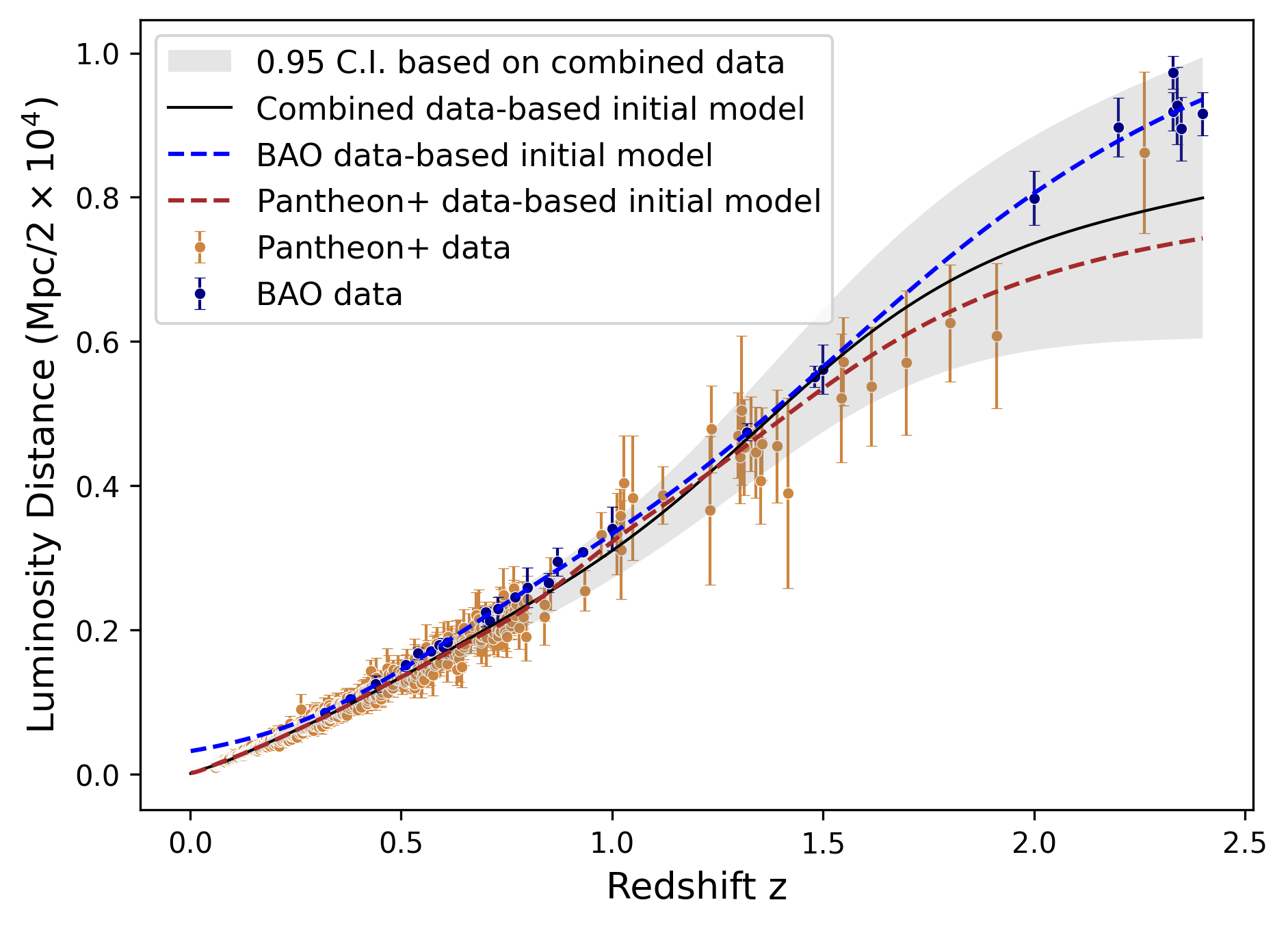}
	\caption{\small Initial models  }
	\label{fig:initial}
\end{subfigure}
  \begin{subfigure}[b]{0.49\textwidth}
\includegraphics[width=\textwidth]{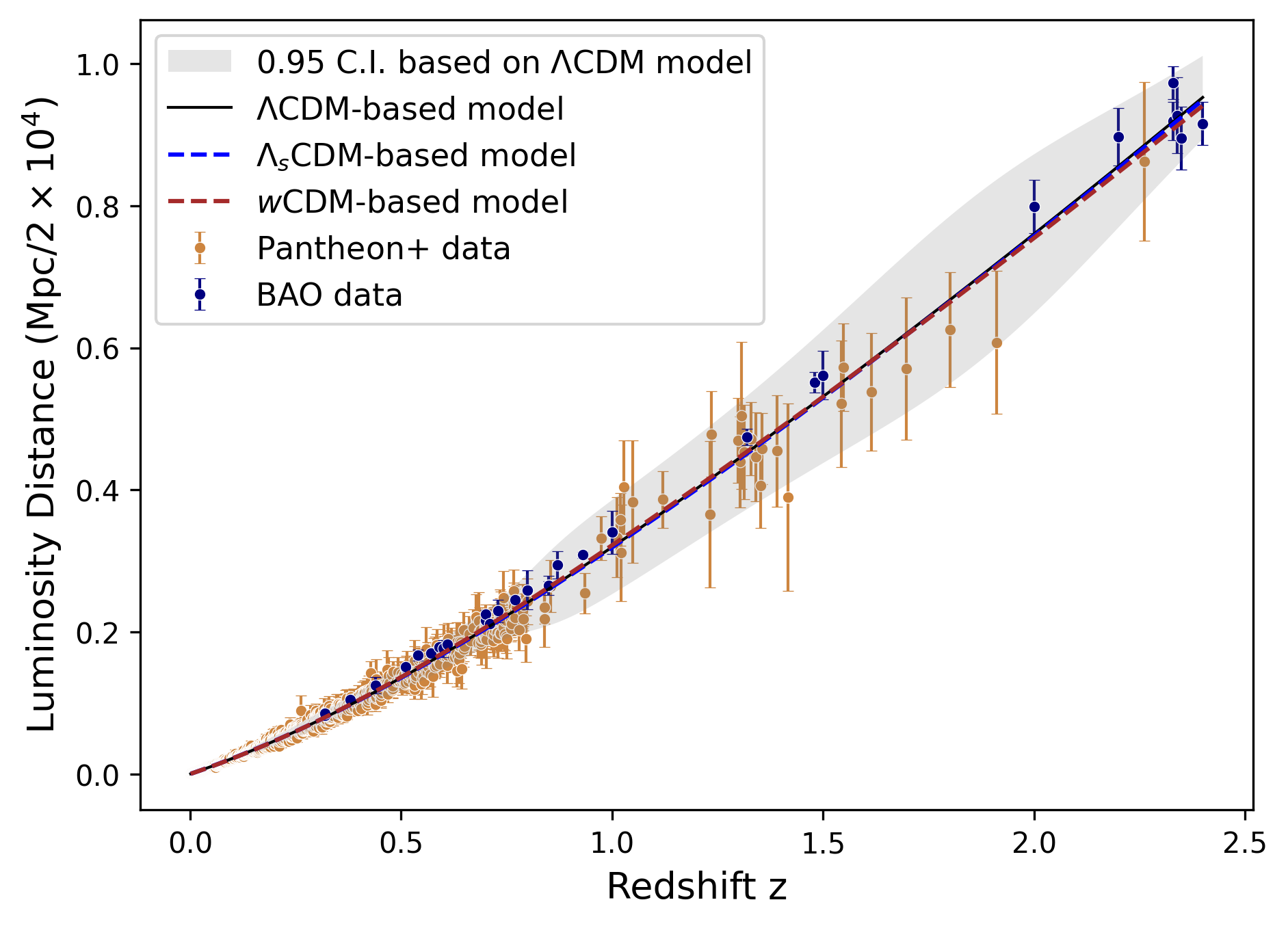}
	\caption{\small Final models }
    \end{subfigure}
\caption{\small Left diagram shows initial models fitted on only Pantheon$+$, BAO datasets and their combination. Right diagram shows the final models fitted on the combined Pantheon$+$ and BAO data. Numerical solutions equipped with the posterior medians (omitted) are all very close to their respective neural network predictions. The purely data-fitted models without PDE constraints suggest that the empirical data trends alone would yield luminosity vs redshift curves of decreasing slope at higher redshift values $z\gtrsim 2$,in contrast to the numerical solutions governed by Friedmann equations.  }
    \label{fig:pb_curves}
\end{figure}

\begin{figure}[h!]   
    \begin{center}		
\includegraphics[width=0.8\textwidth]{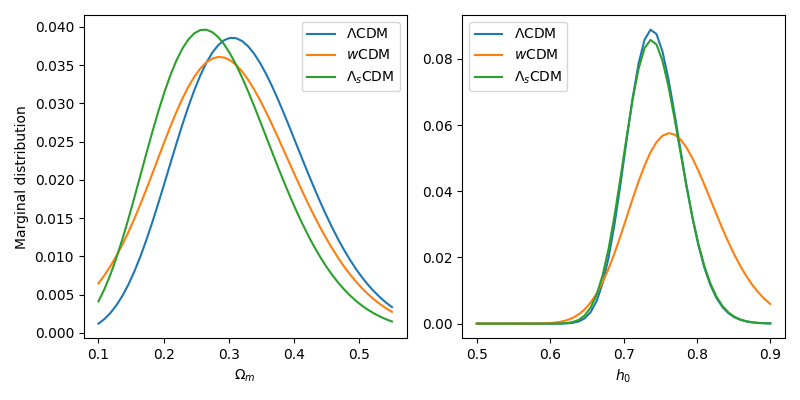}
	\end{center}
    \caption{\small Marginalized posterior distributions for $h_0, \Omega_m$ for all models trained on the combined dataset. Each model yielded lower values of $\Omega_m$ and higher values of $h_0$ than when trained on the individual datasets separately. All three models yielded posterior medians of $h_0, \Omega_m$ that agree within one standard deviation. }
	\label{fig:pb_mar}
\end{figure}

\newpage

\section{Discussion}
\label{sec:conclusion}
In this work, we have applied E-PINN -- a novel variant of Physics-Informed Neural Networks -- to predict cosmological parameters from recent supernovae \cite{Scolnic_2022} and baryon acoustic oscillations (BAO) datasets \cite{Bousis}.  Built upon a hybrid of the principles of Evidential Deep Learning, Physics-Informed Neural Networks and Bayesian Neural Networks, our model enables learning of the posterior distribution of the unknown PDE parameters through standard gradient-descent based training. We also introduced a novel refinement of the original 
E-PINN framework \cite{tan_E,tan_S} that integrates Gaussian Processes into its algorithm, enabling supervised learning of epistemic uncertainty and the construction of prior functions for the model parameters.

Our neural network-based approach introduces a higher degree of model independence relative to standard regression-based statistical analysis, since the perceptron model does not descend from any solutions of some presumed cosmological model while being a fundamental part of the learnable likelihood function. Our approach fundamentally differs from the usual statistical analysis in a few ways: (i)instead of some uniform prior, we use a data-informed prior, constructed to represent an empirical distribution derived from the deviations between the observed data trend and the numerical solution of the presumed PDE; (ii)the loss function that is minimized is generalized from the negative log-likelihood of a Gaussian to a combination of terms (eqn.~\ref{full_loss}) that incorporates both PDE constraints and data loss terms; (iii)Gaussian Process Regression is invoked to supervise learning of epistemic uncertainty; (iv)the surrogate perceptron model extends the standard approach of only using families of PDE solutions for best-fit estimation, enabling the identification of regions where data trends deviate from the presumed PDE descriptions.

With regards to the Hubble tension problem \cite{HTension}, the essential finding of our work is that the posterior distributions for cosmological parameters in various models trained purely on Pantheon$+$ data were found to be largely contained within the $2\sigma$ contours of their counterparts trained on BAO data (Fig. \ref{fig:joint_marginals}).
As tabulated in Table~\ref{tab:results_cosmo}, the $h_0$ values were within about $2 \sigma$ of one another as defined through the marginal distributions in $h_0, \Omega_m$, in contrast to those in \cite{Bousis} exhibiting more than $4\sigma$ tension as inferred from the standard approach of minimizing an appropriate $\chi^2$ function. The normalized residuals (Fig.~\ref{fig:resi_1D}) indicated presence of dataset-dependent systematic effects, where each dataset favors a different set of cosmological parameters -- a trend supportive of some degree of Hubble tension that is consistent with \cite{Bousis}.
In \cite{Bousis}, the best-fit values for the $\Lambda$CDM model associated with the BAO dataset were
$(h_0, \Omega) = (0.67, 0.34)$, while the Pantheon dataset yielded $(h_0, \Omega) = (0.73, 0.33)$. Their posterior distributions showed large deviations as depicted in Fig.~4 of \cite{Bousis} where one can see that their probability contours at 3$\sigma$ do not even overlap. While our framework yielded parameter estimates similar to theirs -- 
$(h_0, \Omega) = (0.68, 0.36)$ based on BAO data and $(h_0, \Omega) = (0.73, 0.36)$ based on Pantheon data, these posterior medians were inferred with larger uncertainties, with posterior distributions that showed much larger degree of overlap in Fig.~\ref{fig:joint_marginals}, compared to Fig.~4 of \cite{Bousis}. One contributing factor can be traced to the nature of the prior density functions that we derived 
for the parameters (see Table ~\ref{tab:prior}). Their uncertainties were slightly larger yet similar in order of magnitude to those characterizing the joint marginal distributions in Fig.~\ref{fig:joint_marginals}. In contrast, typically flat priors are used in the standard analysis, with the eventual parameter uncertainties only dependent on the minimization of the likelihood function in eqn.~\eqref{standard_reg}. Apart from the prior, our framework also assumes a different likelihood function. Instead of eqn.~\eqref{standard_reg} which is expressed in terms of the solutions to the Friedmann equations, we have eqn.~\eqref{resi_exp}. These differences led to our joint marginals as shown in Fig.~\ref{fig:joint_marginals} to deviate from those in the standard analysis of \cite{Bousis}. Notably, our models' inferences of $h_0, \Omega_m$ were equipped with much larger uncertainties (at least twice for $h_0$ and thrice for $\Omega_m$) compared to those arising from the standard analysis
in \cite{Bousis}.

Overall, our simulation results showed that a more data-informed approach can seemingly reduce statistical tensions between models trained separately on Pantheon and BAO data, providing a different measure of the Hubble tension compared to the standard method of minimizing a suitable $\chi^2$ function \cite{Brout_2022,Bousis}. This is fundamentally due to different sets of assumptions underlying the statistics of model parameters, most crucially the choice of prior and form of the likelihood functions, and how they relate to cosmological ones. A caveat is that the sound horizon $r_d$ was taken to be fixed at 147.18 Mpc following Planck18 results \cite{Planck18} when translating $D_M/r_d$ in the BAO dataset of \cite{Bousis} to luminosity distances. While this enables us to compare our results fairly with those of \cite{Bousis}, it should also be interpreted as a limitation, since the tension is sensitive to $r_d$.

All initial models in the absence of PDE constraints arising from presumed cosmological models appeared to suggest that luminosity-redshift curve should flatten out towards higher redshift gradually, in contrast to the numerical solutions for all three cosmological models considered here. It would be interesting to observe if future empirical data from supernovae light curves at high redshift support this trend.
With regards to model selection, we note that the log model evidence appeared to disfavor $\Lambda_s$CDM when the combined Pantheon and BAO data were taken into account, but otherwise showed no other notable model preferences. The $w$CDM model yielded the highest $h_0$ values relative to the two other models irrespective of the dataset used.

An immediate future direction worth pursuing as a follow-up to our work here would be to use a model-independent sound horizon $r_d$ for training models on BAO data, or to lift it to be a learnable parameter. 
A recent analysis \cite{liu_sound} inferred a  value for $r_d \sim 140$ Mpc 
by leveraging time-delay measurements of gravitationally lensed quasars from H0LiCOW collaboration \cite{holicow} in a model-independent approach. Such a value would naively reduce deviations between the models trained separately on Pantheon and BAO datasets, as shown in Fig.~\ref{fig:resi_1D}. As noted in \cite{liu_sound}, future cosmological probes may bring in greater diversity of data sources, such as gravitational wave standard sirens \cite{gw_siren}, with which we can infer the sound horizon and other cosmological parameters. Our data-driven neural network methodology is poised to leverage such increasingly diverse observations to infer parameters with robust data-informed priors. 
More generally, in the aspect of machine learning techniques, we expect our proposed method of synergizing Gaussian Processes with E-PINN to be transferable to other scientific modeling problems, and to be particularly useful for contexts where data is relatively scarce and learning of epistemic uncertainty then becomes crucial to the model training process.  

Another follow-up work would be to explore generalizations of evidential deep learning that can facilitate the learning of known correlations among data points.  
Naively, one can consider the multivariate t-distribution obtained by marginalizing out means and covariances of a multivariate Gaussian with the Normal-Inverse-Wishart prior, but this would imply that the input's dimensionality is fixed to the specific value of the training dataset size, and the number of target variables would be increased by over an order of $10^6$. Another line of approach would be to consider adjusting the relative weightage of data points that are correlated, for example taking some form of weighted mean of a pair of correlated points instead of treating them as separate entities within the sum over all training points. One would need to examine how to take such a weighted mean that is robustly consistent with the off-diagonal elements of the covariance matrix. In the absence of a framework that enables external knowledge of correlation to be used in the supervised learning, a cleaner approach would be to simply use datasets that are not characterized by significant correlations.\footnote{We are grateful to an anonymous referee for raising this point.} For example, in the context of our work, it would thus be interesting to see how our various results change when we use BAO data that is sourced from SDSS\cite{sdss}, DESI\cite{Abdul}, etc. separately instead of Table 1 of \cite{Bousis}.

\section*{Acknowledgments}
I am grateful to Rafe McBeth for many discussions on related topics, including our recent collaborations in \cite{tan_E,tan_S}, and to Phuntsok Tseten for his moral support. I dedicate this work to the loving memory of my aunt, Tan Siew Huan, and my uncle, Tan Hang Song.

\appendix
\renewcommand{\theequation}{A\arabic{equation}}
\setcounter{equation}{0}

\section{Some technical aspects of the loss function}
\label{sec:AppA}
In this Appendix, we present more detailed explanations for certain specific aspects of the loss function.

\subsection{On the t-distribution in EDL}
\label{sec:AppA_t}
Our framework (E-PINN) generates uncertainty estimates for its output by leveraging the principle of \emph{Evidential Deep Learning}(EDL) \cite{amini,sensoy}. The framework of EDL in the context of regression can be summarized as follows. One first considers a probabilistic model where each output neuron is accompanied by another one representing its uncertainty. This pair of neurons can be interpreted as the Gaussian mean and variance for the probabilistic target. We can further assume prior distributions for the mean $\mu$ and variance $\sigma^2$ and integrate $(\mu,\sigma^2)$ out to obtain a marginal distribution that depends on the observed data and the parameters of the prior distribution. Specifically taking the prior to be a normal-inverse-gamma distribution (NIG) with $\mu \sim \mathcal{N}(\gamma, \sigma^2/\nu)$, $\sigma^2 \sim \Gamma^{-1}(\alpha, \beta)$, we obtain the marginal distribution to be
a t-distribution as follows. 
\bea
\label{t_distribution}
&&\iint d\mu d\sigma^2\,\,\,
f_{\mathcal{N}} (L_{obs};\mu, \sigma^2)
f_{NIG}(\mu, \sigma^2; \alpha, \beta, \nu, \gamma) = 
\frac{\Gamma\left( \alpha + \frac{1}{2} \right)}{\Gamma \left( \alpha  \right)  \sqrt{2\pi \beta  (1+\nu)/\nu  }}
\left(
1 + \frac{(L_{obs} - \gamma)^2}{2\beta  (1+\nu)/\nu}
\right)^{-(\alpha+\frac{1}{2})}
\cr
&&f_{\mathcal{N}} (L_{obs}; \mu, \sigma^2) \equiv \frac{1}{\sqrt{2\pi \sigma^2}} \text{exp}\left[
- \frac{(L_{obs} - \mu)^2}{2\sigma^2}
\right],\cr
&&f_{NIG}(\mu, \sigma^2; \alpha, \beta, \nu, \gamma ) \equiv 
\frac{\beta^\alpha \sqrt{\nu}}{\Gamma(\alpha ) \sqrt{2\pi \sigma^2}} \left(  
\frac{1}{\sigma^2}
\right)^{\alpha + 1}
\text{exp}\left[ - \frac{2\beta + \nu (\gamma - \mu)^2}{2\sigma^2} 
\right],
\eea
where $f_{\mathcal{N}}$ denotes the auxiliary Gaussian distribution, $f_{NIG}$ the NIG prior and $L_{obs}$ the observed data. 
Instead of just a single output neuron or a pair representing ($\mu, \sigma^2$), our model has four output neurons ($\alpha, \beta, \nu, \gamma$)
where $\gamma$ represents the mean and $\alpha, \beta, \nu$ are related to the predictive variance $\sigma^2_p$ as follows. 
\be
\label{uncertainties}
\sigma^2_a = \mathbb{E}(\sigma^2) = \frac{\beta}{\alpha - 1},\,\,\,
\sigma^2_e = \text{Var}(\mu)
 = \frac{\beta}{(\alpha - 1)\nu}, \,\,\, \sigma^2_p = \sigma^2_a + \sigma^2_e = \frac{\beta}{\alpha - 1}\left( \frac{1}{\nu} + 1 \right)
\ee
where $\sigma^2_a, \sigma^2_e$
denote the `aleatoric' and `epistemic' uncertainties respectively, with the expectation and variance operators defined with respect to $f_{NIG}$. We note that the aleatoric uncertainty $\sigma^2_a$ is typically interpreted as the uncertainty related to measurement noise while $\sigma^2_e$
represent the uncertainty due to data insufficiency and the underlying model's capacity to represent the observed knowledge (see e.g. \cite{kendall} for a nice discussion). Both types of uncertainties sum up to yield the variance of the t-distribution of 
\eqref{t_pde} (reproduced below for reading convenience).
\be
 P(\mathcal{D} | \mathcal{M}(\vec{w}) )
 =  
\frac{\Gamma\left( \alpha + \frac{1}{2} \right)}{\Gamma \left( \alpha  \right)  \sqrt{2\pi \beta  (1+\nu)/\nu  }}
\left(
1 + \frac{(L_{obs} - \gamma)^2}{2\beta  (1+\nu)/\nu}
\right)^{-(\alpha+\frac{1}{2})}.
\ee
Its negative logarithm is the data loss term $\mathcal{L}_{data}$ discussed in Section~\ref{sec:model}.

\subsection{On uncertainty of $\vec{\Omega}$ and its prior distribution }
\label{sec:AppA_para_prior}

In the Bayesian approach, one should consider specifying a prior density function $\pi (\vec{\Omega})$ for $\vec{\Omega}$. For example, a reasonable choice would be one that is derived from other empirical measurements and inference of $\vec{\Omega}$. Taking into account $\pi(\vec{\Omega})$, the loss function then reads 
\be
\label{loss}
\mathcal{L} = -\log\left[
P(\mathcal{D} | \mathcal{M}(\vec{w}) ) \,
P(\mathcal{M}(\vec{w}) | \vec{\Omega} )  \,\pi(\sigma^2_R; \alpha_r, \beta_r)\,
\pi ( \vec{\Omega} ) \right], 
\ee
in a form interpretable as the negative logarithm of a posterior distribution for $\vec{\Omega}$. The model's weights $\vec{w}$ are latent variables, with model training that is based on minimizing $\mathcal{L}$ equivalent to a 
\emph{maximum a posteriori} estimation. We can compute the uncertainty of $\vec{\Omega}$ as being defined with respect to the posterior density function
\be
\label{posterior}
f_p \left(\vec{\Omega} | \mathcal{D}, \mathcal{M}(\vec{w}) \right)
= \frac{P(\mathcal{M}(\vec{w})|\vec{\Omega}) \pi(\vec{\Omega})}{\int d\vec{\Omega}\,
P(\mathcal{M}(\vec{w})|\vec{\Omega}) \pi(\vec{\Omega})
},
\ee
where we have discarded
$\vec{\Omega}$-independent terms. 
Restoring the input indices, we note that since the data loss term and PDE residual term are products of i.i.d. individual observations, the likelihood function can be expressed as  
\be
P(\mathcal{D}, \vec{w} | 
\vec{\Omega} ) = 
P(\mathcal{D} | \mathcal{M}(\vec{w}) )P(\mathcal{M}(\vec{w}) | \vec{\Omega} )
\equiv
\prod^{N_D}_{j=1}
P(\mathcal{D}_j| \mathcal{M}(\vec{w}), z_j )
\prod^{N_p}_{k=1}
P(\mathcal{M}(\vec{w}) | \vec{\Omega}, z_k ),
\ee
In the following, we will derive a form for the prior $\pi (\vec{\Omega})$ that can be used generally. Let $\mathcal{D}_{\vec{\Omega}}$ denote the finite, discretized domain for the unknown parameters $\vec{\Omega}$. At each point of $\mathcal{D}_{\vec{\Omega}}$, we can evaluate the mean squared deviation between the solution to the PDE characterized by $\vec{\Omega}$  and the Gaussian Process mean $\tilde{\mu}$ in eqn.~\eqref{aux_Gaussian}. 
\be
\label{msd}
F (\vec{\Omega}) = \frac{1}{N_D}\sum^{N_D}_{j=1}
\left(
L_p (z_j; \vec{\Omega}) - 
\tilde{\mu} (z_j)
\right)^2,
\ee
where $L_p (z; \vec{\Omega})$
denotes a numerical solution to the differential equation with parameters $\vec{\Omega}$, and 
$\tilde{\mu} (z_j)$ GP regression model evaluated on $z_j$. 
We assert a Gaussian likelihood based on the mean squared deviation in \eqref{msd} for $\vec{\Omega}$, with the variance parameter being the mean $\overline{F}$ averaged over the domain $\mathcal{D}_{\vec{\Omega}}$. This defines a density function $f$ at each point $\Omega$ of the form 
\be
\label{fund_prior}
f(\vec{\Omega}) = \frac{1}{N} e^{-\frac{F(\vec{\Omega})}{2\overline{F}}},\,\,\, N = \int_{\mathcal{D}_{\vec{\Omega}}}
d\vec{\Omega}\, f(\vec{\Omega}),\,\,\,
\overline{F} \equiv 
\frac{1}{|\mathcal{D}_{\vec{\Omega}}|}
\int_{\mathcal{D}_{\vec{\Omega}}}
d\vec{\Omega}\,
F(\vec{\Omega}),
\ee
where $N$ is the normalization constant and all integrals are  implemented as numerical Riemann over the discretized 
domain $\mathcal{D}_{\vec{\Omega}}$. 
We would like the prior distribution of $\Omega$ to be characterized by the same mode and dispersion scales as the highest density region ( \cite{Hyndman} ) of $f$. At some confidence level, say 68\%, this region is generally a complex subset of the domain $\mathcal{D}_\Omega$. Since our choice of prior distribution affects model training dynamics in the second phase, we adopt a simple Gaussian surrogate distribution for this region, with the means being the modes 
and the standard deviations being those of each marginal distribution. 
\be
\label{prior_formula_1}
\pi (\vec{\Omega}; \vec{\mu}, \Sigma ) 
\sim \frac{1}{\sqrt{\det \Sigma}}
\text{exp}\left[- 
\frac{
\lvert \lvert
\vec{\Omega} - \vec{\mu} \rvert
\rvert^2}
{2\Sigma}
\right],
\ee
where $\Sigma$ is a diagonal covariance matrix of which elements are the variances of the marginal distribution for each component of $\vec{\Omega}$,
while the mean vector $\vec{\mu}$ are the modes of $f(\vec{\Omega})$
\be
\label{stats_prior}
\vec{\mu} = 
\underset{\vec{\Omega}}{\arg \max}\, f(\vec{\Omega}),\,\,\,
\Sigma_{ij} = \delta_{ij} \text{Var} \left[
\int_{\mathcal{D}_{\vec{\Omega}}}
d\Omega_1 \ldots 
d\Omega_{i-1} d\Omega_{i+1}\ldots d\Omega_m
\,\,\,\,\, f(\vec{\Omega}) \right].
\ee
This choice of the prior distribution yields a simple approximation of the highest density region of $f(\vec{\Omega})$ (eqn.\eqref{fund_prior}) which is in turn based on the mean squared deviation between the data-fitted model's curve and the numerical solution equipped with $\vec{\Omega}$, with the dispersion scale in each parameter component $\Omega_k$ set by the variance of its marginal distribution.

\subsection[Determination of $\pi(\sigma^2_R; \alpha_r, \beta_r)$]{Determination of $\pi(\sigma^2_R; \alpha_r, \beta_r)$}
\label{sec:fix_sigmaR}

In this Appendix, we present a detailed discussion of a method that can be used to   set the prior density for $\sigma^2_R$ -- the dynamical, learnable weight for the PDE residual loss term. Its prior density is intended to
guide and regularize the evolution of $\sigma^2_R$ during the gradient descent-based training as the model adapts to both data and PDE constraint. Assuming an inverse-gamma distribution for its form, 
\be
\label{prior_R}
\pi(\sigma^2_R; \alpha_r, \beta_r) =
\frac{\beta_r^{\alpha_r}}{\Gamma (\alpha_r)}
\sigma^{-2(\alpha_r + 1)}_R e^{-\frac{\beta_r}{\sigma^2_R}},
\ee
we pick its hyperparameters $(\alpha_r, \beta_r)$ such that it is consistent with other aspects of our formalism. These parameters are known as the shape and scale factors respectively, in particular leading to the mode and mean values being 
$\frac{\beta_r}{\alpha_r + 1}$ and $\frac{\beta_r}{\alpha_r - 1}$ respectively. Here we restrict ourselves to the case where $\alpha_r >1$ so that the mean is well-defined. 
We pick the initial value of $\sigma^2_R$ ($\equiv \sigma^2_{ini}$) to be the mean. As $\sigma^2_R$ decreases during model training, it approaches the mode of $\pi (\sigma^2_R; \alpha_r, \beta_r)$ at which the derivative with respect to $\sigma^2_R$ vanishes. 
\be
\label{sigma_initial}
\sigma^2_{ini} = \frac{\beta_r}{\alpha_r - 1}, \,\,\,
\sigma^2_{asy} =\frac{\beta_r}{\alpha_r + 1},
\ee
where $\sigma^2_{ini}$ denotes initial value, and $\sigma^2_{asy}$ denotes an asymptotic lower bound at the completion of model training. 
Since the distribution of $\vec{\Omega}$ is defined through eqn.\eqref{posterior}, preceding model training, we would like the initial likelihood function to be close to the prior distribution for $\vec{\Omega}$. This motivates setting $\pi(\sigma^2_R; \alpha_r, \beta_r)$ such that the initial induced statistics of $\vec{\Omega}$ is similar to $\pi (\vec{\Omega}; \vec{\mu}, \Sigma)$. 

To proceed, we first obtain the initial data-fitted model $\mathcal{M}_0$ by training the model using only the EDL loss function augmented with the aleatoric and epistemic loss terms in the first training phase.
\be
\label{pure_data_loss}
\mathcal{L}_{\text{1st phase}} = -\log \left[ 
P(\mathcal{D} |\mathcal{M}(\vec{w}) ) \right] + 
\mathcal{L}_{alea} + \mathcal{L}_{epi},
\ee
where $P(\mathcal{D} |\mathcal{M}(\vec{w}) )$
is defined in eqn.~\eqref{t_distribution}, $\mathcal{L}_{alea}$
is defined in eqn.~\eqref{alea_loss_m} and 
$\mathcal{L}_{epi}$ is defined in eqn.~\eqref{epi_loss_m}.
Thus, this first phase of model training is performed without alluding to any PDE description. Upon convergence, we then obtain $\mathcal{M}_0$ -- a purely data-fitted model. 

We would like the initial induced statistics of $\vec{\Omega}$ in the likelihood function 
$P(\mathcal{M}_0 (\vec{w}^0)| \vec{\Omega} ; \sigma^2)$ to be similar to $\pi(\vec{\Omega}; \vec{\mu}, \Sigma)$, since the latter represents the prior.
Using the Kullback-Leibler divergence as a measure of similarity, we set
\be
\label{sigmaR} 
\frac{\beta_r}{\alpha_r - 1} = 
\underset{\sigma^2}{\arg \min}\,\,\, D_{KL} \left(
P(\mathcal{M}_0 (\vec{w}^0)| \vec{\Omega} ; \sigma^2) \Vert 
\pi(\vec{\Omega}; \vec{\mu},\Sigma )
\right),
\ee
where 
\be
\label{ini_sigmaR}
P(\mathcal{M}_0 (\vec{w}^0)| \vec{\Omega} ; \sigma^2) = 
\frac{\text{exp} \left[- \frac{1}{2\sigma^2}
\sum^{N_D}_{k=1}
\mathcal{R}^2_k \left(
\partial f, f, x_k, \vec{\Omega}
\right) \right]}{\int d\vec{\Omega}\,\,\,\text{exp} \left[- \frac{1}{2\sigma^2}
\sum^{N_D}_{k=1}
\mathcal{R}^2_k \left(
\partial f, f, x_k, \vec{\Omega}
\right) \right]}
\ee
More intuitively, the parameter $\sigma^2_R$ controls the overall scale of the dispersion of each component of $\vec{\Omega}$. The constraint \eqref{sigmaR} sets the initial $\sigma^2_R$ such that the likelihood function 
is initially close (in the sense of KL measure) to the prior function for $\vec{\Omega}$. 

As the model adapts to the PDE residual condition, $\sigma^2_R$ decreases and moves from the mean towards the mode where the derivative with respect to $\sigma^2_R$ vanishes. We would like the minimum uncertainties at this point to be consistent with our model implementation, in particular, the discrete nature of the domains for the parameters $\vec{\Omega}$. These domains are necessarily characterized by finite resolutions. Consider a diagonal multivariate Gaussian distribution $\pi_m (\vec{\Omega}; \vec{\mu}, \Sigma_{min})$ where each standard deviation of $\Sigma_{min}$ is set as the minimal spacing in each parameter's domain. This then yields a natural choice for the mode of $\pi(\sigma^2_R; \alpha_r, \beta_r)$.
\be
\label{sigmaR_mode} \text{mode} \left( \sigma^2_R \right)  = \frac{\beta_r}{\alpha_r + 1} = 
\underset{\sigma^2}{\arg \min}\,\,\, D_{KL} \left(
P(\mathcal{M}_0 (\vec{w}^0)| \vec{\Omega} ; \sigma^2) \Vert 
\pi_m(\vec{\Omega}; \vec{\mu}, \Sigma_{min} )
\right).
\ee
The two KL divergence-minimization equations \eqref{sigmaR} and \eqref{sigmaR_mode} then determine $\alpha_r, \beta_r$ which regularizes the adaptive evolution of the PDE residual loss term weight $\sigma^2_R$.

\section{Some plots of posterior distributions}
\label{sec:AppB}
\renewcommand{\thefigure}{B\arabic{figure}}
\setcounter{figure}{0}
 
\begin{figure}[ht]
    \centering
\includegraphics[width=\linewidth]{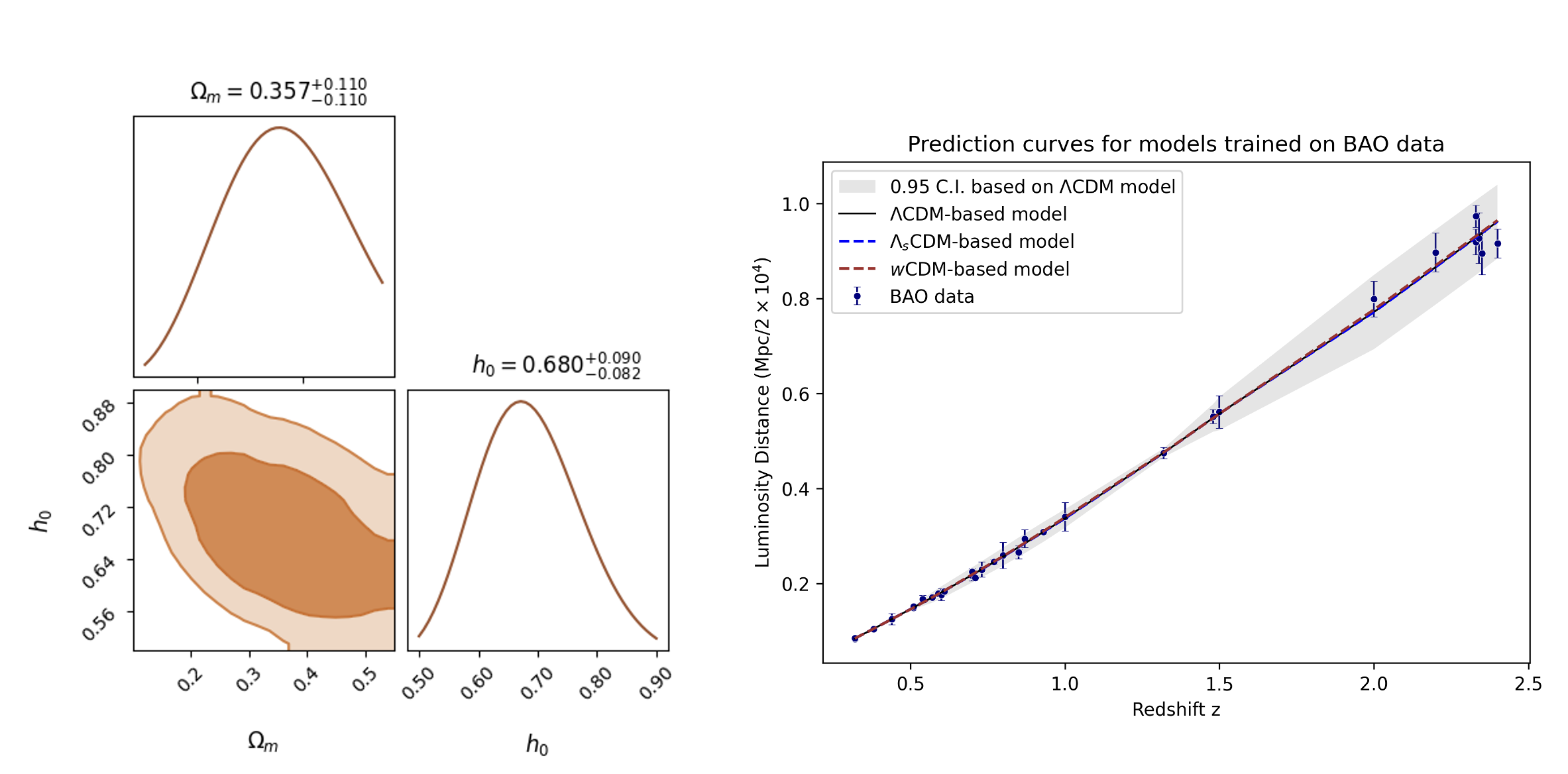}
    \caption{(left) Inferred posterior distribution for $\Lambda$CDM-based model trained purely on BAO data; (right) Prediction curves for all models trained purely on BAO data }
    \label{fig:corner_b1}
\end{figure}

\begin{figure}[h!]
    \centering
    \begin{subfigure}[b]{0.49\textwidth}
        \centering     \includegraphics[width=0.9\linewidth]{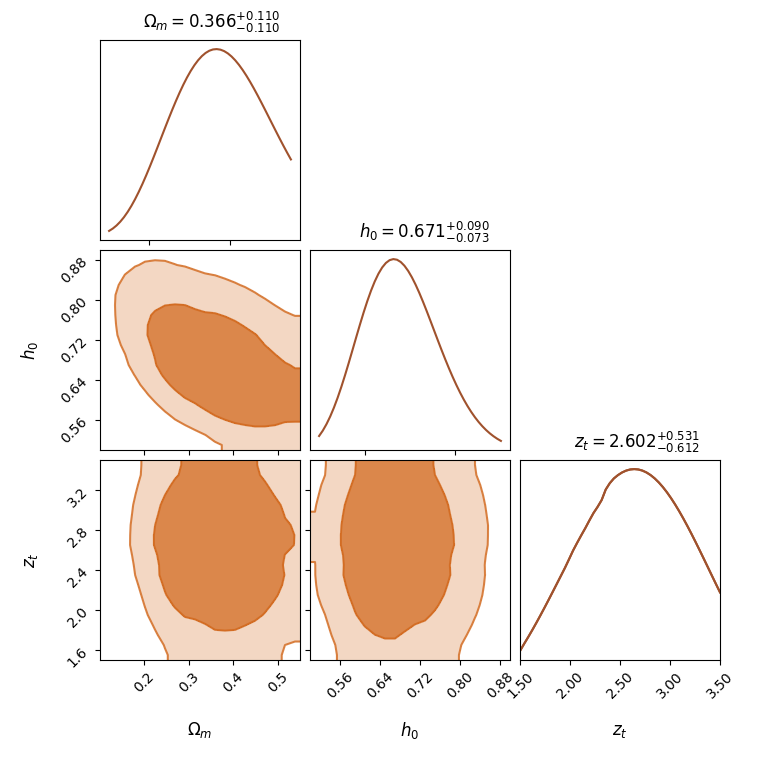}
        \caption{\small $\Lambda_s$CDM-based model (BAO)}
        \label{fig:corner_b_s}
    \end{subfigure}
    \hfill
    \begin{subfigure}[b]{0.49\textwidth}
        \centering      \includegraphics[width=0.9\linewidth]{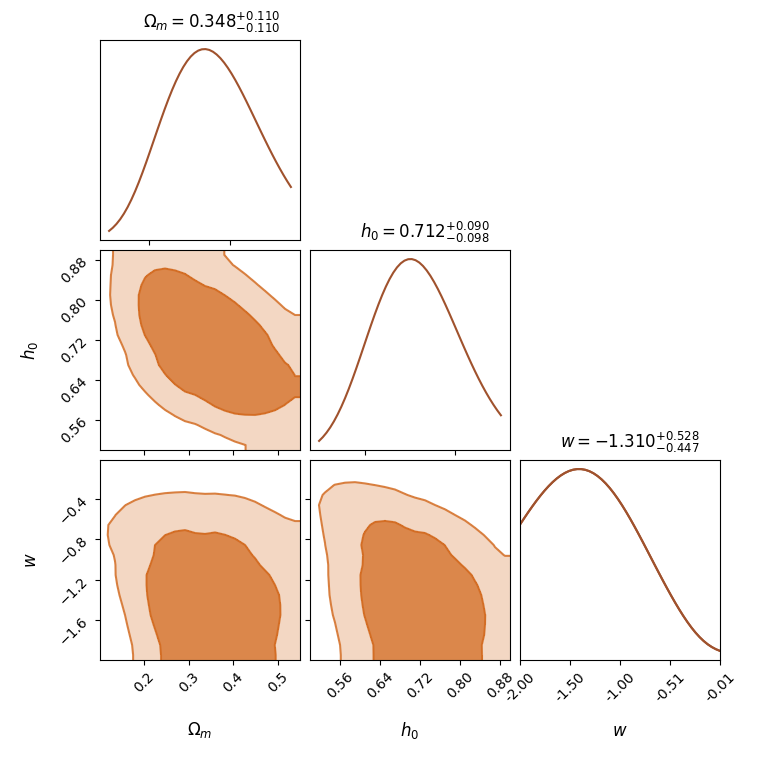}
        \caption{\small $w$CDM-based model (BAO)}
        \label{fig:corner_b_w}
    \end{subfigure}
    \caption{\small Corner plots for the posterior distributions inferred from the $\Lambda_s$CDM-based and $w$CDM-based models trained purely on BAO data.}
\label{fig:corner_b2}
\end{figure}

\begin{figure}[ht]
    \centering
  \includegraphics[width=\linewidth]{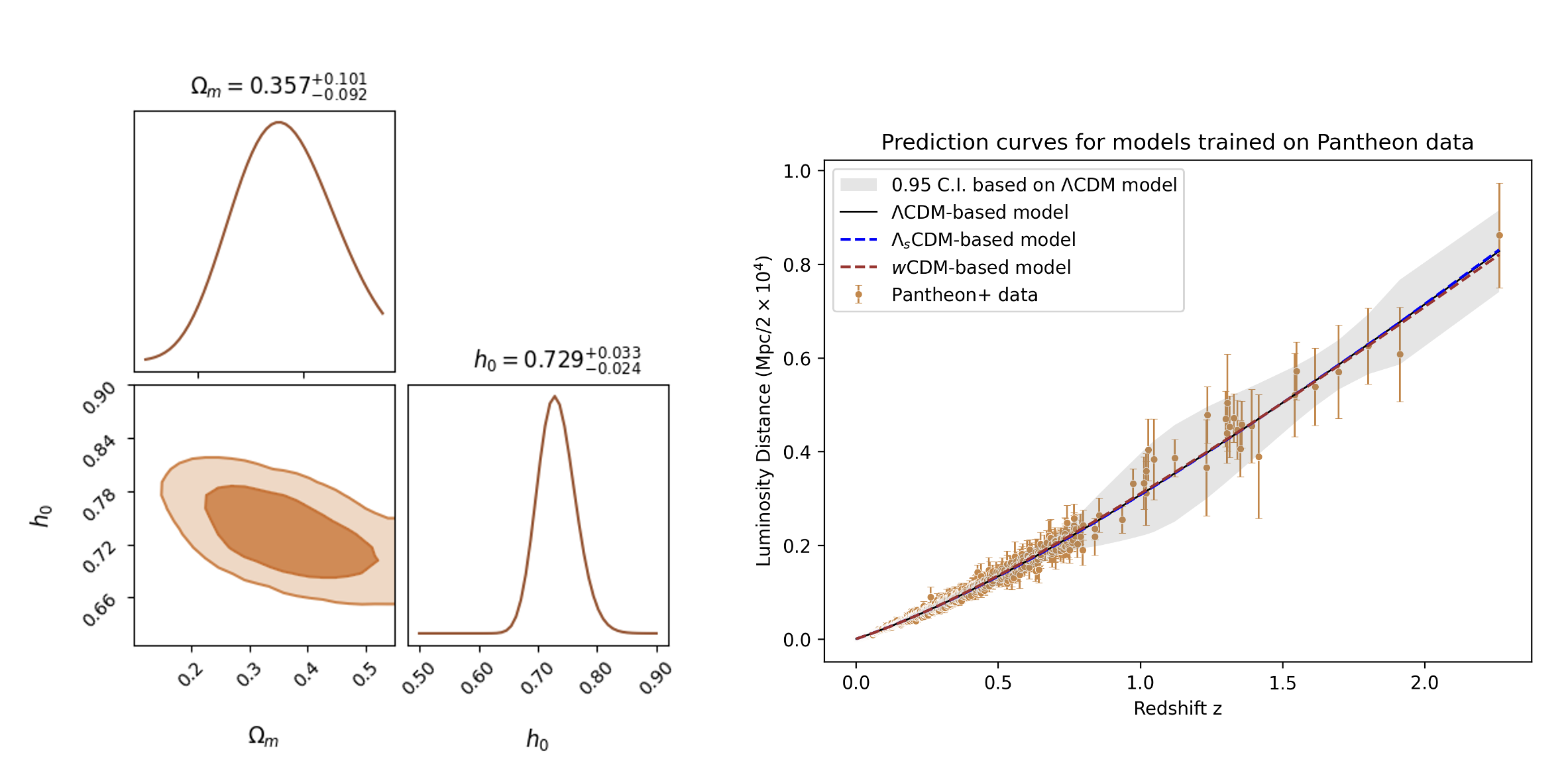}
    \caption{\small (left) Inferred posterior distribution for $\Lambda$CDM-based model trained purely on Pantheon$+$ data; (right) Prediction curves for all models trained purely on Pantheon$+$ data }
    \label{fig:corner_b3}
\end{figure}

\begin{figure}[ht]
    \centering
    \begin{subfigure}[b]{0.49\textwidth}
        \centering     \includegraphics[width=1.0\linewidth]{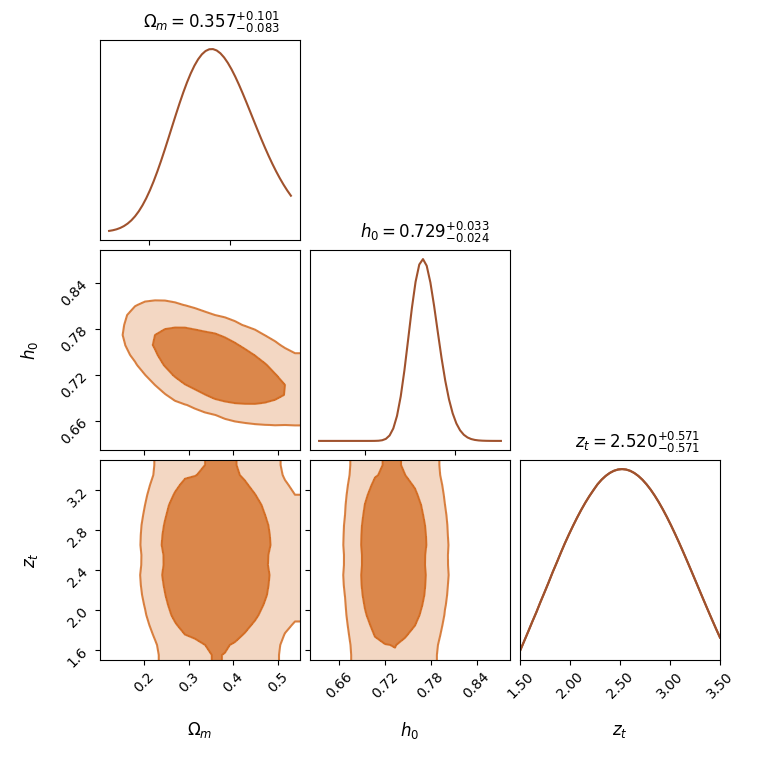}
        \caption{\small $\Lambda_s$CDM-based model (Pantheon$+$)}
        \label{fig:corner_p_s}
    \end{subfigure}
    \hfill
    \begin{subfigure}[b]{0.49\textwidth}
        \centering      \includegraphics[width=\linewidth]{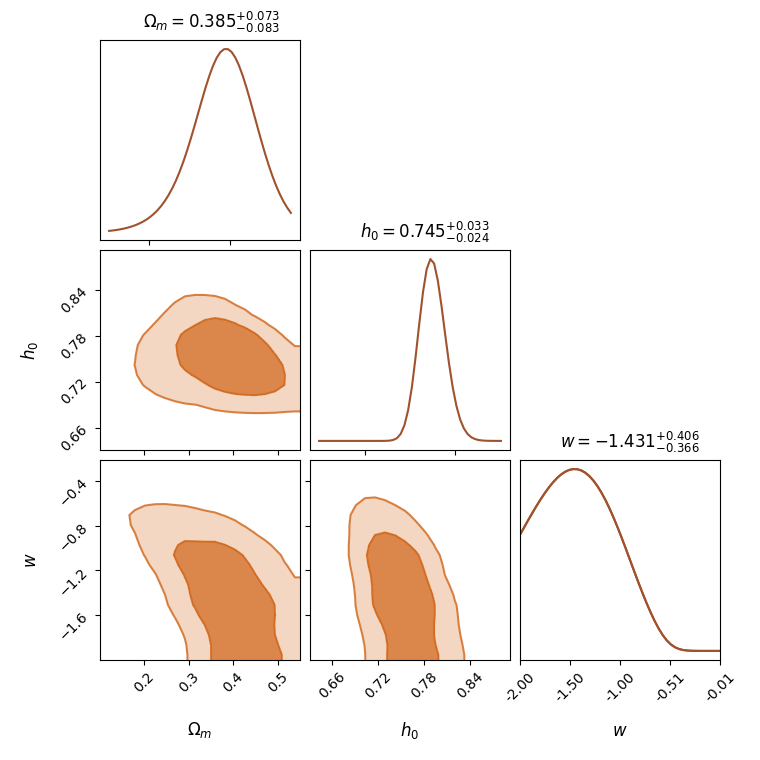}
        \caption{\small $w$CDM-based model (Pantheon$+$)}
        \label{fig:corner_p_w}
    \end{subfigure}
    \caption{\small Corner plots for the posterior distributions inferred from the $\Lambda_s$CDM-based and $w$CDM-based models trained purely on Pantheon$+$ data.}
\label{fig:corner_b4}
\end{figure}

Here, we collect the corner plots for various models trained separately on the BAO (Fig.~\ref{fig:corner_b1},\ref{fig:corner_b2}) and Pantheon (Fig.~\ref{fig:corner_b3},\ref{fig:corner_b4}) datasets.

\newpage 
\clearpage
\bibliography{Project_Epinn_Cosmo}
\bibliographystyle{vancouver}

\end{document}